\documentstyle[useAMS,graphics,epsfig,psfig]{mn2e}
\def\zabs{{\it z}$_{abs}$}
\def\zem{{\it z}$_{em}$~}
\def\lya{Ly$\alpha$ }

\def\h2{H$_2$}
\def\hi{H~{\sc i}~}

\def\cii{C~{\sc ii}~}

\def\kms{km~s$^{-1}$}

\def\nh{$\it{ n_H}$}
\def\N{{\it N}}

\begin{document}
\title[\h2 in DLAs]{Molecular hydrogen in the diffuse interstellar medium at high redshift}
\author[R. Srianand et al. ]
{R. Srianand$^{1}$, 
 G. Shaw$^{2}$,
G. J. Ferland$^{2}$,
P. Petitjean$^{3,4}$,
C. Ledoux$^{5}$
\\
${}^1$ IUCAA, Postbag 4, Ganeshkhind, Pune 411007, India;
anand@iucaa.ernet.in \\
${}^2$ Department of Physics and Astronomy, University of
Kentucky, 177 Chem.-Phys. Building, Lexington, KY 40506, USA;\\
  gshaw@pa.uky.edu, gary@pa.uky.edu\\
${}^3$ Institut d'Astrophysique de Paris -- CNRS, 98bis Boulevard Arago, F-75014 
Paris, France; petitjean@iap.fr\\
${}^4$ LERMA, Observatoire de Paris, 61 Avenue de l'Observatoire, F-75014, Paris, France\\
${}^5$ European Southern Observatory, 
Alonso de C\'ordova 3107, Casilla 19001, Vitacura, Santiago, Chile; 
cledoux@eso.org
}
\date{Typeset \today ; Received / Accepted}
\maketitle
\begin{abstract}
%
The physical conditions within damped \lya systems (DLAs) can reveal 
the star formation history, determine the chemical composition of the 
associated ISM, and hence document the first steps in the formation of present day 
galaxies. Here we present calculations that self-consistently determine the 
gas ionization, level populations (atomic fine-structure levels and
rotational levels of \h2), grain physics, and chemistry. 
We show that 
for a low-density gas (\nh$\le$ 0.1 cm$^{-3}$) the meta-galactic 
UV background due to quasars is sufficient to maintain  \h2 
column densities below the detection limit (i.e \N(\h2)$\le10^{14}$ cm$^{-2}$) 
irrespective of the metallicity and dust content in the gas. Such a gas will have
a 21 cm spin temperature in excess of 7000 K and very low C~{\sc i}
and C~{\sc ii$^*$} column densities for H~{\sc i} column densities
typically observed 50 per cent in DLAs. 
\par 
We show that the observed properties of the $\sim15$ per cent of the 
DLAs that do show detectable \h2 absorption cannot be reproduced 
with only the quasar dominated meta-galactic UV radiation field. 
Gas with higher densities (\nh$\ge10$ cm$^{-3}$), a moderate 
radiation field (flux density one to ten times that
of the background radiation of the Galactic ISM), the observed range of metallicity and
dust-to-gas ratio reproduce all the observed properties of the 
DLAs that show \h2 absorption lines. 
%
This favors the presence of ongoing star formation in DLAs with \h2.

\par
The absence of detectable \h2 and C~{\sc i} absorption in a large 
fraction of DLAs can be explained if they originate either 
in a low-density gas or in a high-density gas with a large 
ambient radiation field. The absence of 21 cm absorption and
C~{\sc ii$^*$} absorption will be consistent with 
the first possibility. The presence of 21 cm absorption and 
strong C~{\sc ii$^*$} without \h2 and C~{\sc i} absorption will 
suggest the second alternative. 
The \N(Al~{\sc ii})/\N(Al~{\sc iii}) ratio can be used to understand the 
physical properties when only C~{\sc ii$^*$} absorption is present.
We find \nh~in components that show C~{\sc ii$^*$} (without \h2) is less
than that typically inferred from the components with \h2 absorption.
We also calculate the column density of various atoms in
the excited fine-structure levels.
The expected column densities
of O~{\sc i$^*$}, O~{\sc i$^{**}$}, and Si~{\sc ii$^*$} in a 
high-density cold gas is in the range of 10$^{11}-10^{12}$ cm$^{-2}$ for 
log~N(H~{\sc i})$\ge20$ and the observed range of metallicities.
It will be possible to confirm whether DLAs that do not show \h2
originate predominantly
in a high-density gas by detecting these lines in 
very high S/N ratio spectra. 

\end{abstract}


\section{Introduction}

Damped \lya systems seen in QSO spectra are characterized by 
very large H~{\sc i} column densities, \N(H~{\sc i})$\ga 10^{20}$ 
cm$^{-2}$, that are similar to those observed through gas-rich spiral 
galaxies. The importance of DLAs in the paradigm of hierarchical 
structure formation can be assessed from the fact that the mass 
density of baryonic matter in DLAs at \zabs $\sim 3$ is similar 
to that of stars at present epochs (Wolfe 1995). Studies of \lya and 
UV continuum emission from galaxies associated with DLAs usually 
reveal star formation rates(SFR) (or upper limits) of a few 
M$_\odot$ yr$^{-1}$ (Fynbo et al. 1999; Bunker et al. 1999; 
Kulkarni et al. 2001; M\"{o}ller et al. 2002 and 2004;
Weatherley et al. 2005). 
%
The metallicity and depletion factor in DLAs are usually estimated 
from \N(Zn~{\sc ii})/\N(H~{\sc i}) and \N(Fe~{\sc ii})/\N(Zn~{\sc ii}) 
respectively (Lu et al. 1996; Pettini et al. 1997; 
Prochaska \& Wolfe 2002; Ledoux et al. 2002a and Khare et al 2005).
The inferred metallicities typically vary between 
log {\it Z} = $-$2.5 to 0 for $2\le$ \zabs $\le 3$ 
with a median of $\simeq-1.3$ (Ledoux et al. 2003). 
The measured depletions range between 0 to $-1.6$ with a median value 
of $-0.3$. If dust content is defined as 
$\kappa~=~10^{\rm [Zn/H]}~(1-10^{\rm [Fe/Zn]})$, then the median dust 
content in a typical DLA is $\kappa=0.07$. This is  less than
10 per cent of what 
is seen in the Galactic ISM for a similar neutral hydrogen column 
density.
If the physical conditions in DLAs are similar to those in our Galaxy or
the Magellanic clouds then \h2 molecules should be detectable. 
There have been very few detections of \h2 in DLAs (with 
${\rm 3\times10^{14}\le \N(H_2)(cm^{-2})\le 3\times 10^{19}}$) despite 
extensive searches (Ge \& Bechtold 1999; Srianand, Petitjean \& Ledoux 2000;
Petitjean, Srianand \& Ledoux 2000; Ledoux, Srianand \& Petitjean 2002b;
Levshakov et al. 2002; Ledoux, Petitjean, \& Srianand 2003; Reimers et al. 2003).
Roughly 80 per cent of DLAs do not have detectable \h2 
(with ${\rm \N(H_2)}\le10^{14}$ cm$^{-2}$). 

The physical conditions in the \hi gas can be probed by using the 
fine-structure absorption lines produced by the excited atomic 
species and the 21 cm transition. 
Apart from a few rare cases (for example Srianand \& Petitjean 2001), 
C~{\sc i} is detected only in systems that show \h2. The derived
total hydrogen density (\nh) based on the fine-structure level 
populations of the heavy elements are usually
large ($\ge 20 $ cm$^{-3}$) (Ledoux et al. 2002b;
Srianand et al. 2005; Wolfe et al. 2004).
Like C~{\sc i}, C~{\sc ii$^*$} absorption is detected in every case 
where one detects H$_2$.  
However, it has also been seen with lower column densities
in a considerable fraction of DLAs without \h2 (Wolfe et al. 2003a;
Srianand et al. 2005, Wolfe et al. 2004).

The search for 21 cm absorption in DLAs at \zabs$\ge2$ have 
mostly resulted in 
null detections with typical spin temperatures $\ge10^3$ K
(Table 3 of Kanekar \& Chengalur (2003) and Table 1 of
Curran et al. (2005) for a summary of
all the available observations). There are 8 DLAs
at \zabs$\ge$ 1.9 for which redshifted 21 cm observations are
available. Redshifted 21 cm absorption is detected for only
two systems (\zabs = 1.944 toward PKS 1157$+$014 (Wolfe et al. 1981) and 
\zabs = 2.0394 toward PKS 0458$-$02 (Briggs et al. 1989)).
The measured spin temperatures are 865$\pm$190 K and 384$\pm$100 K. 
However, none of these systems show detectable \h2 (Ledoux et al. 
2003; Ge \& Bechtold 1997). The DLA toward  PKS 1157$+$014 is
special as the QSO shows broad absorption lines and the \zabs~ of
the DLA is close to the \zem of the QSO. The physical conditions
in this system may not be a good representative of the general DLA
populations.
Interestingly, \h2 is seen  
at \zabs = 2.8110  toward PKS 0528-2505, while no
21 cm absorption is detected in this system (Carilli et al. 1996;
Srianand \& Petitjean 1998). 
The lower limit on the spin
temperature derived for this system is 710 K. However, the
excitation temperature derived from \h2 rotational levels is
$\le 200$ K (Srianand \& Petitjean 1998; Srianand et al. 2005). 
This system is also
special since \zabs$>$\zem. The radiation field of the QSO
has a much stronger influence on the physical conditions in
this DLA (Srianand \& Petitjean 1998).
The upper limits on the spin temperature derived 
for the rest of the systems are higher than 1000 K.
The \h2 content of these systems is not known.

%
Even though various properties of DLAs, listed above,  have been
investigated in detail (Matteucci et al. 1997;
Prantzos \& Boissier 2000; Howk \& Sembach 1999; Izotov et al.
2001; Vladilo, 2001; Liszt 2002; Hirashita et al. 2003; Wolfe et al. 2003a,b;
Calura et al. 2003; Wolfe et al. 2004), very few attempts have 
been made to investigate all of them in a single unified calculation. That is 
the main motivation of this paper.

\section{Calculations }

The main aim of our study is to investigate the physical conditions
in high-{\it z} DLAs. In particular, our goal is to
understand the equilibrium abundance of \h2, the excitations of
\h2~and atomic fine-structure levels, the ionization state
of the gas, and the 21 cm optical depth, self-consistently. 
In the Galactic interstellar medium (ISM) \h2 is usually detected 
either in a diffuse medium irradiated by the
UV background radiation field or in high-density photo-dissociation 
regions (PDRs) near OB stars. One can anticipate this 
to also be the case in DLAs. At high redshift the 
diffuse UV background from QSOs will be an additional source
of UV radiation.

Recently there have been three attempts to model \h2 in DLAs.
Liszt (2002) uses two phase models similar to Wolfire et al. (1995) 
for this purpose. The models consider dust-free gas, so 
only the slow gas-phase H$^{-}$ + H $\rightarrow$ \h2 + e formation process is important.
The second attempt  by Hirashita et al. (2003) models  
DLAs as large protogalactic disks. The density and temperature in the 
gas are determined by "non-linear hydrodynamic" effects. 
The radiation field is assumed to have a negligible contribution
to the temperature of the gas and is used only for destroying the
\h2~molecules. 
Their model provides insight into the spatial distribution of \h2.  
The third attempt by Hirashita \& Ferrara (2005) uses
simple \h2 equilibrium formation models to study \h2 in DLAs. 
Unlike Liszt (2002), no attempt is made 
to model ionization conditions of the gas and 
excitations of the fine-structure lines in the later two
studies.
The main aim of our paper is to use full self-consistent 
numerical calculations
to understand (i) physical conditions in DLAs with \h2 detections,
(ii) the reason for the lack of \h2 in most of the DLAs
(iii) the origin of C~{\sc ii$^*$} absorption frequently seen
in DLAs and (iii) the absence of detectable 21 cm absorption
at high redshifts (i.e z$\ge2$). 

%
\par
The availability of good quality observational data 
allows us to estimate the metallicity, depletion,
\h2 abundance, \h2 excitation, and populations of
fine-structure levels in C~{\sc i} and C~{\sc ii}
(Ledoux et al. 2003; Srianand et al. 2005). 
One can hope to build more realistic models with all these in hand.
This forms the prime motivation of this work.
We study the ionization state, chemical history, and temperature 
of the gas using version 96 of Cloudy, last described by Ferland et al.(1998), and available at {\bf http://www.nublado.org}. 
The details of the new \h2 model are given in Shaw et al. 
(2005; hereafter S05).
A comparison between predictions of our code and several independent 
calculations of PDRs can be found 
at {\bf http://hera.ph1.uni-koeln.de/$\sim$roellig}. A direct application of a
PDR is given in Abel et al. (2004; hereafter A04). 
The  calculations presented here 
take into account various heating (e.g. dust photo-electric
heating cosmic ray heating etc) and cooling processes
(for details see Ferland et al. 1998 and S05).

%
\subsection{The micro-physics of grains :}

We use the improved grain physics and molecular network
as described in van Hoof et al. (2004), Ferland et al. (1994; 2002), 
and A04. The grain size distribution is resolved, and the formalism
described by Weingartner \& Draine (2001a; van Hoof et al. 2004) 
is used to determine the grain charge, temperature, 
photoelectric heating, and drift velocity
self-consistently with the local radiation field and gas 
temperature. The extinction curves of grains in DLAs are not 
well known. We use a grain size distribution that fits the 
extinction curve of the  diffuse interstellar medium with 
R$_{\rm V}$ = 3.1 (Table~1 of Weingartner \& Draine (2001b)). 
 We emphasize that the physical treatment of grains, and their 
effects on the surrounding gas, is fully self-consistent, and 
does not rely on general fitting formulae, such as those in 
Bakes \& Tielens (1994).  The grain charge and temperature are 
determined by the local gas conditions (mainly the electron 
density and temperature) and radiation field (including the
attenuated incident continuum and emission by the surrounding 
gas, mainly \lya).  The result is a grain temperature and 
charge that depends on grain size and material.  
The temperature then determines the rate of \h2 formation
on grain surfaces - we adopt the temperature-material-dependent 
rates given by Cazaux \& Tielens (2002).  The charge 
establishes the floating potential of
the grain, which then sets the grain photoelectric heating rate.

\subsection{ Molecular hydrogen :}
The detailed treatment of the micro-physics of \h2 is described in S05. Here we will briefly mention some of those 
processes. 

\par
Generally, \h2 forms via grain catalysis in a dusty cold-neutral 
gas. We use the total formation rate given by Cazaux \& Tielens
(2002) along with the size and temperature resolved grain 
distribution described in van Hoof et al.(2004).
This is an exothermic process and \h2 is formed in exited 
vibrotational levels, a process referred to as formation pumping.
The results presented below use the state-specific formation 
distribution function given by Takahashi (2001) and 
Takahashi \& Uehara (2001).
By contrast, in an equipartition distribution function
 1/3 of the binding energy is statistically distributed 
as internal excitation (Black \& van Dishoeck 1987).
Both produce an ortho-para-ratio (OPR) that is nearly 3. 
    
\par
\h2 is formed by associative detachment from 
H$^-$ in a dust-free gas. 
This process is important in the clouds with lower
dust content considered below. This is also an exothermic process and we use the state-specific formation distribution function given by
Launay et al. (1991).

\par
\h2 is destroyed mainly via the Solomon process, in which the
\h2 molecule is irradiated by far UV (912\AA $<$ $\lambda$ $<$ 1200\AA) radiation and is excited to higher electronic states.  
Approximately 10 per cent of 
these electronically excited \h2 decay to the ground state continuum and are dissociated.  The other 90 per cent populate the higher
vibrotational levels of the ground electronic
state. These cascade to lower vibrotational levels producing
infrared emission lines, referred to as Solomon pumping. Formation pumping on grains is only 10 percent as effective as 
Solomon pumping when Solomon destruction
is balanced by grain formation. Thus, the \h2 populations are non-thermal if the electronic lines are optically thin and the Solomon
process is dominant (hereafter
referred to as the optically thin case), while the \h2 level populations may be in LTE at the local gas kinetic temperature if the electronic lines are optically thick and
the Solomon process is slow (hereafter referred to as the optically thick case). 
\par
Radiative decays between ortho and para states are not possible because of the different nuclear spin. However, exchange collisions
with H, H$^+$, H$_{3}$$^+$, and interactions on grain surfaces (below a certain critical temperature) can cause an ortho-para
conversion. 
The column density ratio of {\it J}=1 and {\it J}=0 traces the kinetic temperature in a collisionally dominated gas but may fail to do so in a
Solomon-process dominated region (Abgrall et al. 1992; Sternberg \& Neufeld 1999; Roy, Chengalur \& Srianand (2005)). 
 
\par
The formation rate on dust has the largest uncertainty among the many processes
considered in our calculations. There are significant variations in this rate even in the case of the Galactic ISM 
(Browning et al. 2003). There are also substantial differences between
collisional rates of \h2 computed by different groups at low temperatures (S05). 
These uncertainties should be kept in mind while comparing our results with observations.

\subsection{H~{\sc i} spin temperature:}
It is commonly assumed that the N(H~{\sc i}) spin temperature, 
{\it T$_s$}, is equal to the gas kinetic temperature.
The optical depth of the 21 cm
transition is proportional to N(H$^0$) / {\it T$_s$}, 
so is sensitive to the inverse of the gas kinetic temperature.
The mean value of {\it T$_s$} we
report here is T$_K$ with weighting by N(H$^0$) / T. 
A separate paper (Shaw et al. 2005) 
discusses our treatment of {\it T$_s$}, and
relationships between T, {\it T$_s$}, and \h2 temperature indicators, in detail.
In DLAs, {\it T$_s$}
is usually estimated using the integrated optical depth $\tau_v$ in the
21 cm absorption line and \N(H~{\sc i}) measured from \lya using
\begin{equation}
T_s = { N(H~{\sc i})f_c\over 1.823\times10^{18}\tau_v}.
\label{eqn1}
\end{equation}
Here, $f_c$ is the fraction of the background radio source covered
by the absorbing gas. Thus, low 21 cm optical
depths could either mean high {\it T$_s$} or low $f_c$. 
Even for $f_c = 1$, the derived
temperature will be high if the mean N(H~{\sc i}) covering
the extended radio source is lower than the measured 
one along the line-of-sight toward the optical point source
(Wolfe et al. 2003a).
Thus, observations
will constrain either the physical conditions or the projected
H~{\sc i} surface density distribution of the absorbing gas. 

\subsection{Fine-structure level population:}
The ionization potential of C$^0$ is close to the energy of the 
photons responsible for the H$_2$ electronic band transitions.
So, C~{\sc i} lines may be sensitive to the conditions in the \hi $-$ \h2 transition region.  
The fine-structure level populations of C~{\sc i} are
sensitive to the gas pressure and the IR radiation field. Thus, 
populations of the excited fine-structure levels of C~{\sc i} 
allow an independent probe of
quantities that control the physical conditions in the
\hi $-$ H$_2$ transition region and the temperature of the cosmic microwave 
background (CMBR) (Srianand et al. 2000; Silva \& Viegus 2002).
The column densities of excited levels within the ground term of 
C~{\sc i}, \cii, Si~{\sc ii}, and O~{\sc i} are all calculated as part 
of the gas cooling function.
All excitation and deexcitation processes, collisions, line trapping,
destruction by background opacities, and pumping by the external
continuum, are included. At  high {\it z} the IR pumping is predominantly
by CMBR pumping, although the diffuse IR radiation from  
grains also contributes.

\subsection{Cloud geometry:}
\begin{figure*}
\centerline{\epsfig{file=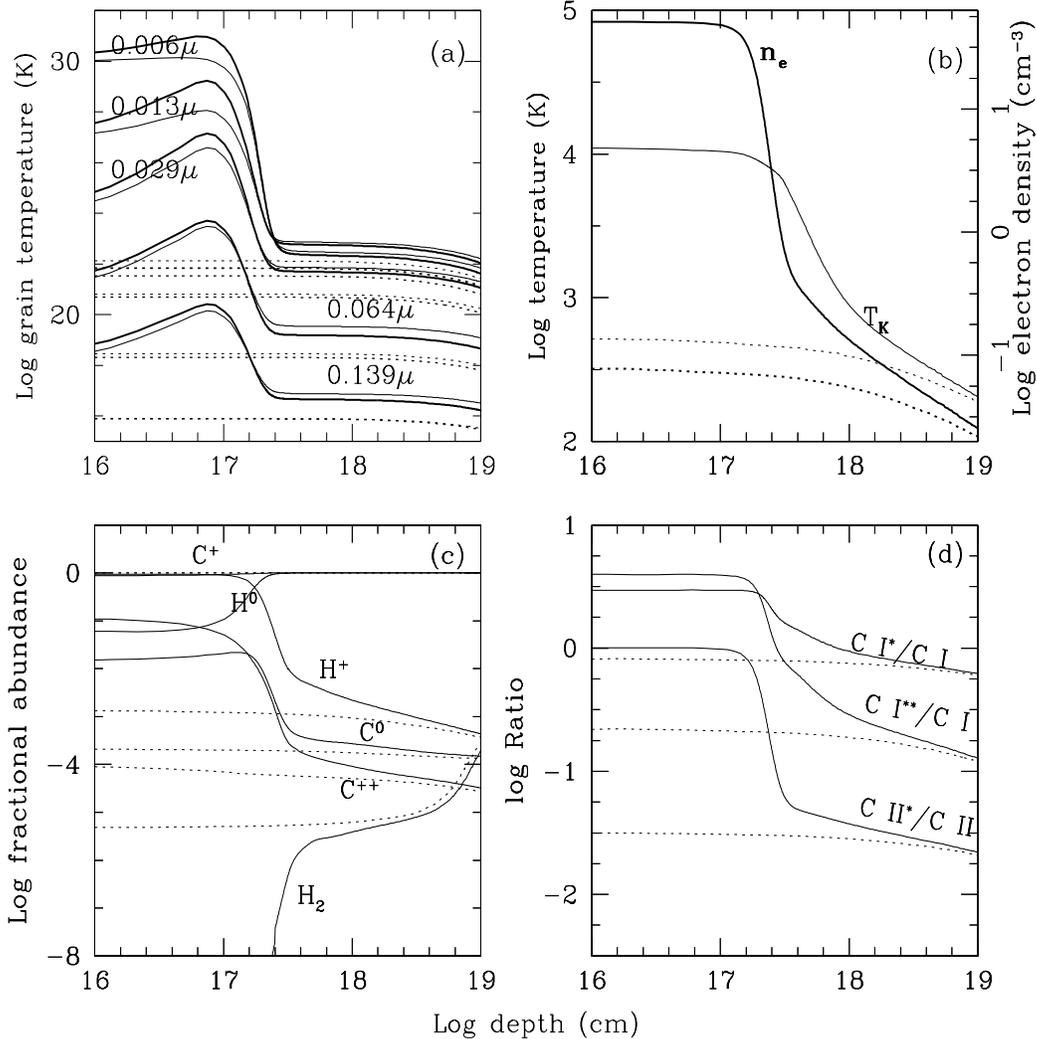,width=14cm,height=15cm}}
\label{depth}
\caption{
{\bf Pedagogical example:} Various physical quantities are shown as a function of depth in a
cloud irradiated by the stellar and diffuse continua with log
\N(H~{\sc i}) = 20.7. The metallicity is 0.1 {\it Z}$_\odot$, \nh= 50 cm$^{-3}$, and log $\kappa$ is $-1.39$ (corresponds to a dust to
metal ratio  0.4 times that seen in the Galactic ISM). 
In these panels the short-dashed lines are for the diffuse case. In panel (a) thick and thin curves are for Silicate 
and Graphite  grains respectively. The short-dashed curves are for models with
diffuse continua. The assumed radiation field is more or less 
4 times that of Galactic mean UV field. In panel (b) the thick curves represent electron density. Panels (c) and (d)
show the ionization and fine-structure excitations as a function of
depth in the cloud. 
}
\end{figure*}

We envision the region where the absorption lines form as a layer or cloud of ISM gas exposed to several radiation fields.
In keeping with much of the PDR literature, we assume a plane parallel geometry (Tielens \& Hollenbach 1985; Draine \& Bertoldi 1996;
Wolfire et al. 1995, 2003; Wolfe et al. 2003a,b). We further assume that the gas has constant density for simplicity. 

\par
In the absence of ongoing star formation, the meta-galactic background UV radiation field, dominated by QSOs (Haardt \& Madau 1996), will
determine the physical conditions and abundance of \h2. If there is ongoing star formation then a locally-produced stellar radiation field will
also contribute.  OB stars are very short lived, and so do not move far from their birthplace before dying. So, newly formed OB
stars will be close to their parent molecular cloud throughout most of their lives. This geometry is assumed in the PDR
references cited above.
\par
Our main goal is to understand the physical conditions in the
components with \h2 and C~{\sc i}.  Only the total  H~{\sc i} column density is measurable in DLAs and the
 H~{\sc i} column density in the \h2 component is generally unknown.
So, we consider clouds with three values of \N(H~{\sc i}); 10$^{19}$, 10$^{20}$, and 10$^{21}$ cm$^{-2}$. We assume the gas metallicity to be 0.1 {\it Z}$_\odot$ and vary 
the dust-to-metal ratio in the range 0.001 to 0.1 
(this corresponds to a range in $\kappa$ (as defined in Section 1)
of $10^{-4}$ to $10^{-2}$) of the galactic ISM.
\par
We consider three ionizing continua; the meta-galactic radiation
field at {\it z} = 2, the direct radiation field from an O star, and an O star continuum that has been attenuated by intervening
absorption.
The first mimics the case in which there is no
{\it in situ} star formation. The second is observed in galactic star-forming regions - the OB stars are close to the
molecular cloud and an H II region lies between them. This will be called the stellar case below. The third would be similar to a
diffuse ISM exposed to the galactic background starlight, and will be called the diffuse case from now on.
\par
Following the general practice in the PDR literature, we define the intensity
of the incident UV radiation field using a dimensionless constant
$\chi$ (as defined by Draine \& Bertoldi 1996),
\begin{equation}
\chi = {\int_{912\AA}^{1110\AA}h^{-1} \lambda u_\lambda~ d \lambda\over
1.22\times10^7}
\label{eqchi}
\end{equation}
Here, $\lambda u_\lambda$ is the energy density (ergs cm$^{-3}$) of 
photons and $\chi$ = 1 for the Galactic UV field defined by
Habing (1968). Thus $\chi$ provides the UV field strength
in the units of Galactic mean UV field.
\par
We use the observed metallicity, depletion,
\h2 abundance, and fine-structure excitations of
C~{\sc i}, C~{\sc ii}, and \N(C~{\sc i})/\N(C~{\sc ii}) to
constrain either the particle density or the intensity of the
radiation field. The 21 cm spin-temperature and the level populations
of \h2 are used for consistency checks.

\subsubsection{Ionization and thermal structure :}
{
In this sub-section we demonstrate the need for
a composite self-consistent simulation of the gas in order to 
deduce the correct physical conditions using a pedagogical example.
In Fig.~\ref{depth} we show the ionization and thermal structure 
of a cloud irradiated by stellar and diffuse continua.

\par
Panel (a) plots the temperature of graphite and silicate grains 
(for the range of sizes considered in our calculations) as a function of
depth from the illuminated side. All the calculations we present in
this work use the self-consistently estimated grain temperature
for a range of grain sizes
 which are important for different processes 
such as photoelectric heating and formation of \h2 on the grain surfaces. 

\par
The kinetic ($T_K$) and the electron 
density ($n_e$) are plotted in panel (b). 
%
{\bf In this work we present the H~{\sc i} weighted harmonic mean 
kinetic temperature as spin temperature ($T_S$)}. Detail 
investigations of  relationship between $T_S$ and $T_K$ 
under different 
physical conditions is described in
Shaw et al.(2005). 

\par
Panel (c) plots the densities of \h2, 
C$^0$, C$^+$, and C$^{++}$ as a function of cloud depth.
The ratio of carbon fine-structure levels are shown in panel (d). 
The electron temperature is high ($\sim$10$^4$ K) and the 
electron density is nearly equal to the H$^+$ density at the 
illuminated side of the gas for the stellar continua 
(panel b, solid line).  
A hydrogen ionization front occurs at a depth of 
2.5$\times$10$^{18}$ cm, where {\it T} and {\it n$_e$} fall.  
Across the PDR, {\it T} ranges between 300 $-$ 800K
and electrons are mainly donated by C$^+$. 
The short-dashed lines show the results for the diffuse case.  
There is no H~{\sc ii} region, and so the entire cloud is a PDR. 
The physical conditions are nearly constant across the cloud, 
which does not have enough grain opacity to attenuate the 
incident continuum significantly.

\par
The behavior of C$^0$ in the case of the stellar case is 
as follows. In photoionization equilibrium, {\it n}(C$^0$) is
$\propto$ $n_e$ {\it n}(C$^+$) $\alpha_{rec}$ $\propto$ $n_e$ {\it n}(C$^+$) $T^{-0.6}$ where $\alpha_{rec}$ is the 
recombination co-efficient. Here $n_e$ decreases by three dex across
the ionization front, however the electron temperature
changes by less than two dex. This leads to two orders of 
magnitude decrease in {\it n}(C$^0$) across the ionization 
front (see upper panel). As the ionization potential of C$^0$ is very
close to the energy of photons that are responsible for the
excitations of electronic states in \h2, one expects both 
C~{\sc i} and \h2 to originate from the same part of the cloud.
This happens for the diffuse case. But in the
case of the stellar case, a considerable fraction of
C~{\sc i} originates in warmer gas that does not possess
\h2. 

\par
The predicted ratio of fine-structure level populations depends on the 
nature of the radiation field. The ratios are  constant 
across the cloud in the case of the diffuse case. However, they
strongly depend on the radiation field for the stellar case.

\par

}

\section{Ionization by the meta-galactic UV background:}
\begin{figure*}
\centerline{\epsfig{file=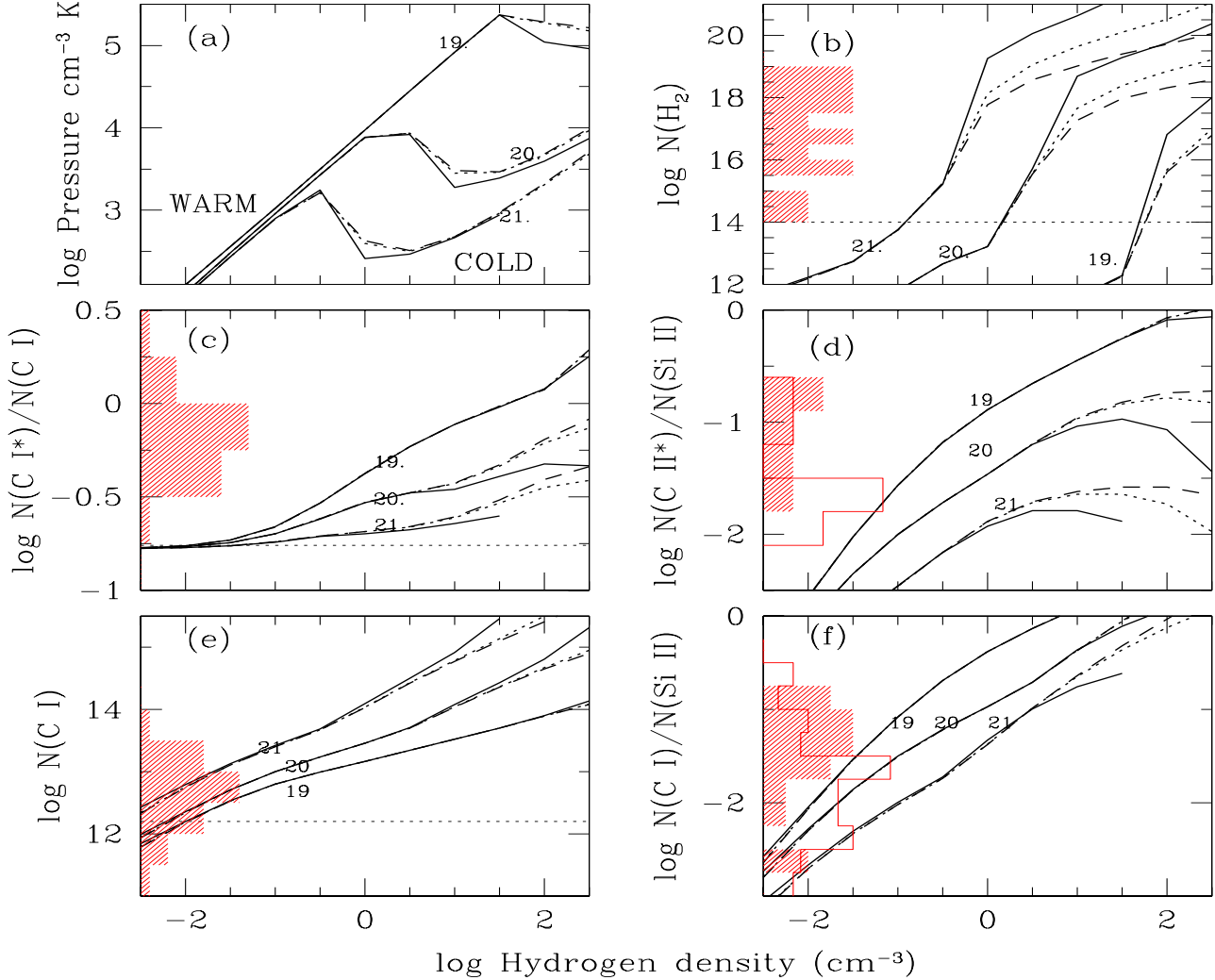,width=18cm,height=15cm}}
\caption {The results for various constant density clouds ionized by the 
meta-galactic UV background given by Haardt \& Madau (1996). 
The results with continuous, short-dashed and long-dashed are for $\kappa$ = 0.01, 0.001 and 0.0001 respectively. 
Panel (a) plots the mean gas pressure 
as a function of hydrogen density. The density-ranges of the  
cold and warm neutral medium are marked in this panel.
The numbers near each line give the assumed
log \N(H~{\sc i}). 
This panel is useful to identify the warm and
cold components of the stable two-phase medium under 
pressure equilibrium.   
The shaded histogram
gives the observed distribution in the systems
with \h2 detections. The non-shaded histogram in panel
(d) gives the results for the systems without \h2.
In panel (f) the non-shaded histogram provides the distribution
of upper limits.
%
The observational data used are mainly from Ledoux et al. 
(2003) and Srianand et al. (2005). The horizontal short-dashed lines 
in panels (b), (e) and (f) are typical detection limits 
obtained in echelle spectra.
%
The horizontal short-dashed line in panel (c) gives the expected 
value of the ratio \N(C~{\sc i}$^*$)/\N(C~{\sc i}) when 
CMBR at {\it z} = 2 is the only source of excitation.  
\label{fig1}}

\end{figure*}

First we consider the case in which the only 
available radiation field is the 
meta-galactic UV background. 
We use the QSO dominated 
meta-galactic UV radiation field (Bgr) computed by 
Haardt \& Madau (1996) and the 
cosmic microwave background radiation (CMBR) at {\it z} = 2
(assumed to be a black body with {\it T} = 8.1 K).  
The UV flux density in
the Bgr at energies below 1 Ryd at {\it z} = 2 is roughly 
two orders of magnitude
lower than the current mean Galactic UV field (i.e $\chi = 1.44\times10^{-2}$
for the Bgr at z = 2).
Some of the results from our calculations are presented
in Figs.~\ref{fig1} and \ref{fig2}.

\subsection{Gas pressure:}
Panel (a) of Fig.~\ref{fig1} plots the mean pressure of 
the H~{\sc i} gas as a function of the total hydrogen density 
(\nh) for three different values of \N(H~{\sc i}). 
Although ionization and thermal gradients exist between the H~{\sc ii} and
H~{\sc i} regions, the H~{\sc i} gas is fairly isothermal. Thus, we use the neutral hydrogen weighted 
mean temperature to estimate the pressure.  
The continuous, short-dashed and long-dashed curves in all these
panels are for dust to metallicity 0.1, 0.01, and 0.001 
(i.e $\kappa$ = 0.01, 0.001 and 0.0001) respectively .  
The regions of thermal stability occur for $d(log P)/d(log n)$$>0$.
The warm neutral medium (WNM) 
and cold neutral medium (CNM) of the two-phase medium 
are shown for reference. As an example, 
gas with 0.3 $\le$ \nh $({\rm cm}^{-3})\le 1$ is thermally unstable for N(H~{\sc i})$\simeq10^{21}~{\rm cm}^{-2}$.
The gas will
be in the stable WNM phase for 0.03$\le$ \nh $({\rm cm}^{-3})\le 0.3$ and 
in the stable CNM phase for 1$\le$ \nh $({\rm cm}^{-3})\le 30$.
For reference, Fig. 6 of Wolfire et al. (1995) shows a similar phase diagram for various metallicities and dust content.
The allowed minimum and 
maximum pressure in the two-phase medium are higher
in our case than in the typical galactic ISM for a given column density, mainly
because of the low metallicity and low dust-to-gas ratio
(see also Petitjean et al. (1992), Lizst (2002), Wolfe et al. (2003a,b);
Wolfe et al. (2004)).
The main motivation for plotting the phase 
diagram from our calculations is to have a rough idea of the 
nature of the gas
at different densities and to compare 
our work with published models. It is worth keeping in mind the fact
that ISM is more complex than different phases in pressure
equilibrium with one another. For example magnetic field,
if present in DLAs, can provide confinement even if there
is no thermal pressure equilibrium between different phases.
Thus, we do not make any serious attempt to model the DLAs 
as two-phase systems.

\subsection{\h2 abundance:}

In this section, we compare the predicted and measured \N(\h2)
values (Ledoux et al. 2003) to determine the
physical conditions in clouds both with and without observed \h2.

\subsubsection{Systems without \h2 detections:}
First we consider the cases  where \h2 is not detected.
The  predicted \h2 column densities as 
a function of hydrogen density (\nh) are shown in panel 
(b) in Fig.~\ref{fig1}. The horizontal
short-dashed line gives the typical detection limit achieved
in \h2 surveys (i.e., \N(H$_2$)$ = 10^{14}~{\rm cm}^{-2}$).
The observed  \N(\h2) is distributed uniformly 
between  10$^{14}$ and 10$^{19}$ cm$^{-2}$ (The histogram
in the left hand side in panel (b)).  

\par
From this figure it is clear that for a given 
\N(H~{\sc i}), the column density of \h2 is independent 
of $\kappa$ when the density is low and the gas is mainly the WNM.
This is mainly because in the low-density, high-temperature gas, 
H$_2$ is formed predominantly through the H$^-$ process due to the low dust-to-gas ratio.
It is also clear that 
the hydrogen
density has to be higher than 0.1, 1.0, 30
cm$^{-3}$ for \N(H~{\sc i}) = 10$^{21}$, 10$^{20}$, and
10$^{19}$ cm$^{-2}$ respectively in order to detect \h2. In the presence of an additional local radiation
field (perhaps generated due to {\it in situ} star formation) these
critical densities will be larger.
{ Thus, if the gas in DLAs is mainly a stable 
WNM in ionization equilibrium with the Bgr, then the equilibrium 
\h2 column density will be below the detection limit.}
This inference is independent of $\kappa$ since the H$^-$ process dominates the \h2 formation at low
densities.

\par
Fig.~\ref{fig1} suggests that if DLAs have a 
thermally stable CNM  then the equilibrium abundance of \h2 
is high enough for the molecule to be easily detectable 
whenever $\kappa$ is greater than 0.0001 or
dust to metal ratio grater than 0.001 
(The long-dashed curves in panel (b)). 
The \h2 formation time-scale, which can be long, does not 
affect this result.
A typical time-scale for forming \h2 with molecular fraction 
$f_{\rm H_2}$ is $\sim$$f_{\rm H_2}$/2$Rn$(H$^0$). Here {\it n}(H$^0$) denotes the atomic hydrogen density.   
According to Jura (1975) R$\simeq 3\times10^{-17}$
cm$^{+3}$ s$^{-1}$ in the Galactic interstellar medium. 
Scaling this value by $\kappa$ we have,
\begin{equation}
{\rm
t = {5.025 \times 10^8~f_{H_2}\over\kappa~n_H}~ yrs.
}
\end{equation}
Assuming that all the hydrogen is H$^0$, we find a typical \h2 formation 
time-scale of $\sim 5 \times 10^5$ yrs with $\kappa$ = 0.0001
and ${\rm f_{H_2} = 10^{-6}}$ for \nh~=10 cm$^{-3}$. 
The age of the
cloud has to be less than 10$^{5}$ yrs for us not to detect \h2 with \N(H$_2$)$\le10^{14}$ cm$^{-2}$ and $\kappa=0.0001$.  
The hydrodynamical time-scales or pressure readjustment
time-scales 
in the cold gas are usually larger than this value
(Hirashita et al. 2003) due to the low sound speeds. Hence the typical age of the
clouds is expected to be larger.
Thus 
\h2 should be detectable in a CNM with $\kappa$ more than 0.0001 
in the absence of any additional local radiation field.  

\subsubsection{Systems with \h2 detections:}

Now we focus on the systems with detectable \h2. 
Ledoux et al. (2003) have shown that these systems  usually have 
a high metallicity and dust content, 
i.e; {\it }Z$\ge0.1${\it Z}$_\odot$ and log~$\kappa\ge-2$. 
If the gas originates from a stable CNM,  
our calculations predict ${\rm \N(H_2)\ge10^{19}}$ 
cm$^{-2}$ (Panel (b) of Fig.~\ref{fig1}) for \N(H~{\sc i}) $\ge 10^{20}$ cm$^{-2}$. 
The observed \N(\h2) is 
always smaller. Interestingly, the observed 
\N(H$_2$) values  with $\kappa\ge0.01$ are only reproduced 
in a very narrow density range, one that is usually thermally 
unstable in the standard two-phase models. This signifies that, for  
a uniformly distributed range of densities, 
a cloud with a random choice of \nh,~ with \N(H~{\sc i}) in the range of 10$^{20}$ 
to 10$^{21}$ cm$^{-2}$, and with $\kappa\ge0.01$, will have either no 
detectable \h2 or \N(H$_2$) $\ge 10^{19}$ cm$^{-2}$.
The probability of detecting \h2 in the observed column density range of 10$^{16}$
to 10$^{19}$ cm$^{-2}$ 
is very low.  {Thus, in order to understand the relatively low column 
densities of \h2 in DLAs with clear detections, we need 
either the presence of an additional
radiation field or for the cloud to be too young to produce \h2.}
The existence of a local radiation field with intensity of the order or higher
then the Galactic mean field has been suggested by various
authors while discussing \h2 in individual DLAs
(Black et al. 1987; Ge \& Bechtold 1997; Srianand \& Petitjean
1998; Petitjean et al. (2000;2002); Ge, Bechtold \& Kulkarni 2001;
Ledoux et al. 2002b, Levshakov et al. 2002; Reimers et al. 2003). 

\subsection{Fine-structure excitations of C~{\sc i} and C~{\sc ii}:}
In this section, we compare the column densities predicted for 
C~{\sc i} and C~{\sc ii} 
fine-structure excitations with the observations. Many of these results will not agree leading us to conclude that an
additional source of radiation must exist in these systems. The time-scales to establish the ionization and
populations within the fine-structure levels of neutral atoms are faster than the \h2 formation time-scale, so this provides an
indicator which should be in steady state.

\subsubsection{C~{\sc i} absorption: detectability and degree
of ionization}
The \N(C~{\sc i})/\N(C~{\sc ii}) ratio is a good tracer of the flux of 
photons driving the Solomon process in the cold neutral gas 
where H$_2$ forms since the ionization potential of C$^0$ overlaps with the \h2 electronic bands. However, \N(C~{\sc ii}) cannot
be accurately determined in DLAs since the C~{\sc ii}$\lambda1334$ line is 
usually saturated.
Unlike C~{\sc ii}, the column densities of Si~{\sc ii} and  S~{\sc ii}
(which usually trace the same region as C~{\sc ii} (see Fig.~\ref{depth})) 
are accurately measured using transitions  with low oscillator 
strengths. Si and S are usually
not highly depleted in DLAs (but see Petitjean et al. 2002 for a 
unique counter example).   
\N(Si~{\sc ii}) can be used as
a proxy to estimate \N(C~{\sc ii}) (Srianand et al. 2000) by assuming the solar abundance of Si/C. 
Now we compare the predicted \N(C~{\sc i}) and \N(C~{\sc i})/\N(Si~{\sc ii})  
with the observations.
We plot \N(C~{\sc i}) and \N(C~{\sc i})/\N(Si~{\sc ii})
as a function of \nh~respectively in panels (e) and (f) of Fig.~\ref{fig1}.  
The predicted values of \N(C~{\sc i}) are much closer 
to the detection limit in a very low-density gas (\nh$\le0.01$ cm$^{-3}$). As previously noted, our calculations are 
performed with Z = 0.1Z$_\odot$, and the systems that do not 
show C~{\sc i} absorption tend to have metallicity in the 
range of ${\rm 0.01 Z_\odot\le Z\le0.1Z_\odot}$
(Srianand et al. 2005). Thus, the absence of 
C~{\sc i} is consistent with a cloud having 
\nh$\le0.01$ cm$^{-3}$, for the metallicity and \N(H~{\sc i})
typically measured in these systems. 
Such a cloud will have log~\N(C~{\sc i})/\N(Si~{\sc ii})
$\le-1.5$ (Panel (f) of Fig.~\ref{fig1}). This is
consistent with the measured upper limits, shown as the non-shaded
histogram in panel (f), in most of the systems without detected \h2.
But \N(C~{\sc i}) is more than an order of magnitude larger than  
the typical detection limit for \nh$\ge1$ cm$^{-3}$. Thus,  
C~{\sc i} should be detectable 
in a high-density gas, even for low metallicity, in the absence of any 
internal radiation field. The predicted values of 
\N(C~{\sc i})/\N(Si~{\sc ii}) in the high-density gas is 
usually higher than the observed upper limits (see
Liszt 2002 and Wolfe et al. 2003a).

\par
Next we concentrate on systems with detectable
C~{\sc i} absorption. As noted in section 1, apart from only 
one case (\zabs = 2.139 toward Tol~1037$-$270), all the 
C~{\sc i} detections 
are from DLAs that also have \h2.
These systems have metallicities higher than we assume.
Based on the detection of \h2, we expect
these systems to also have a higher density.
Dense clouds (\nh$\ge0.1$ cm$^{-3}$) produce a higher value
of \N(C~{\sc i}) than observed.
The measured ratio of  \N(C~{\sc i})/\N(Si~{\sc ii}) in
these components is much less than our predictions  
for a high-density gas. Clearly higher radiation field
is needed to produce N(C~{\sc i})/N(Si~{\sc ii}) as
measured in these systems.
%
\par
The fine-structure populations of C~{\sc i} and C~{\sc ii}
can also test the high density requirement for components with \h2 detections. This will be discussed in the next subsection.

\subsubsection{C~{\sc i} fine-structure excitation:}
Here we use the observed \N(C~{\sc i}$^*$)/\N(C~{\sc i}) ratio to constrain \nh~in the components with \h2.  
This ratio is regularly used to 
trace the pressure in a neutral gas (Jenkins \& Tripp 2001; 
Srianand et al. 2005). We plot  
\N(C~{\sc i$^*$})/\N(C~{\sc i}) as a function of hydrogen density in panel (c) of Fig.~\ref{fig1}.  
The dotted line gives the expected value of 
the ratio if CMBR pumping at {\it z} = 2 is the only source 
of C~{\sc i} fine-structure excitation. 
The observed values of \N(C~{\sc i}$^*$)/\N(C~{\sc i})
(the histogram in the left hand side) are much higher than this,
suggesting that collisions and 
UV pumping also contribute to the excitation. 
Most of the observed ratios are consistent with the 
predictions for  \nh~$\ge~10~{\rm cm}^{-3}$ for the considered range of \N(H~{\sc i}) and $\kappa$. 
{As noted above, for such a high-density gas,
our calculations predict \N(\h2) and \N(C~{\sc i})/\N(Si~{\sc ii})  
higher than observed. We show below that the presence
of an additional radiation field can reduce both of these.}

\subsubsection{Excitations of C~{\sc ii} fine-structure level:}

Like C~{\sc i}, C~{\sc ii$^*$} is always detected whenever \h2
is present in DLAs. However, it is also seen in a considerable
fraction of DLAs without C~{\sc i} and \h2 (Wolfe et al. 2003a;
Srianand et al. 2005). The observed column density of C~{\sc ii$^*$} 
can directly give cooling rate (when the optical depth of
[C~{\sc ii}]$\lambda158$ line is negligible) which can be
used to constrain the star-formation rate once
the physical conditions in th gas is known (Wolfe et al.
2003a; 2003b; 2004).
C~{\sc ii} 
can originate from the CNM, WNM, and an ionized gas 
(i.e., H~{\sc ii} regions). Collisions with 
atoms are important for exciting C~{\sc i$^*$}, while electron 
collisions are important for the excitation of 
C~{\sc ii$^*$}. Thus, C~{\sc ii$^*$} is expected to be detectable
in systems with a warm and/or ionized gas even if the high-density 
CNM is absent (see Lehner et al (2004) and Srianand et al. (2005)).

C~{\sc ii$^*$} is invariably detected 
in all the systems  with log \N(H~{\sc i})$\ge 21$ and 
log~{\it Z}$\ge-0.03${\it Z}$_\odot$ irrespective of the absorption redshift.
%
%
The observed column 
density is in the range 12.7$\le$log \N(C~{\sc ii}$^*$)$\le$14.0 for systems at 1.5$\le$\zabs$\le$2.5. 
The typical upper limit is \N(C~{\sc ii}) $\le$ 10$^{13}$ cm$^{-2}$ (Srianand et al. 2005).  
The calculated  
\N(C~{\sc ii$^*$})/\N(Si~{\sc ii}) ratio is shown in panels (d) 
of Fig.~\ref{fig1}. 
The shaded and non-shaded 
observed histograms in these panels are for the systems
with and without \h2 detection respectively. {The observed range of
\N(C~{\sc ii$^*$})/\N(Si~{\sc ii}) in systems with 
\h2 detections suggests they originate in
a high-density gas.
This is consistent with our conclusion based on \h2
and \N(C~{\sc i}$^*$)/\N(C~{\sc i})}.

The observed \N(C~{\sc ii$^*$})/\N(Si~{\sc ii}) ratio 
tends to be smaller in systems without \h2. 
As we have mentioned above, most of these systems have total 
\N(H~{\sc i}) higher than 10$^{21}$ cm$^{-2}$. {Thus, 
the systems that
show C~{\sc ii$^*$} 
will
originate in clouds with \nh$\ge$0.1 cm$^{-3}$.
Thus, we will require a radiation field in excess
of the Bgr to suppress C~{\sc i} and \h2 in a high-density gas}.

\subsection{H~{\sc i} spin temperature:}
\begin{figure}
\centerline{\epsfig{file=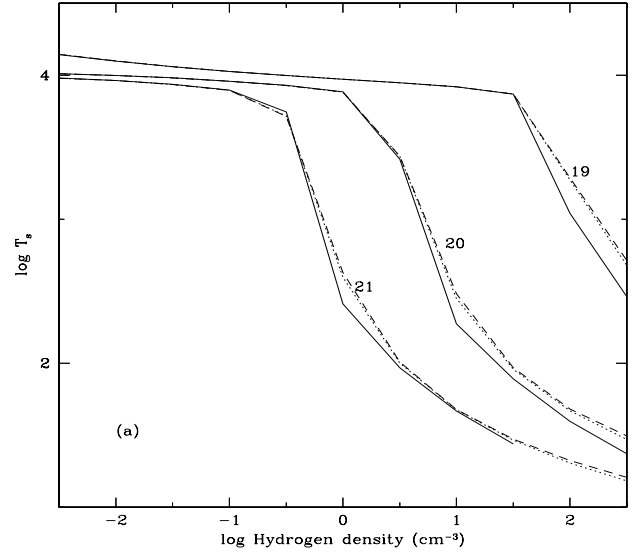,width=9cm,height=8cm}}
\caption {The calculated H~{\sc i} spin temperature is shown as a function
of density with the meta-galactic
UV background dominated by the QSOs at $z\simeq2$. 
The results are presented for three different column densities of \N(H~{\sc i})(10$^{21}$, 10$^{19}$, 10$^{20}$, 10$^{21}$ cm$^{-2}$)
and three different values of $\kappa$ (continuous, short-dashed and long-dashed
curves are for $\kappa$ = 0.01, 0.001, and 0.0001 respectively).
\label{fig2}}
\end{figure}

The thermal state of  H~{\sc i} gas can be probed with   
the 21 cm spin-temperature ($T_s$). The harmonic weighted mean temperature, a proxy for $T_s$,  
%
is shown in panel (a) of 
Fig.~\ref{fig2}. It is clear that if DLAs originate in a 
low-density WNM gas, then the spin temperature will be 
$\simeq 8000$ K. Thus, systems with no \h2, C~{\sc i}, C~{\sc ii$^*$}, 
and 21 cm absorption are consistent with a low-density 
WNM in radiative equilibrium with the Bgr. 
The predicted spin temperatures are usually less than 100 K
for clouds with N(H~{\sc i})$\ge 10^{20}$ cm$^{-2}$ in the CNM.
Thus, we expect all DLAs to show detectable 21 cm absorption if they originate in a high-density CNM gas 
which covers the background radio source.
Unlike C~{\sc i} or \h2, the presence of an additional radiation
field with {\it h}$\nu$ $\le$ 13.6 eV may not reduce the 21 cm optical depth because it can not ionize hydrogen.
The absence of 21 cm absorption in most DLAs
suggests that they originate in a low-density
warm medium. Our calculations also suggest that H~{\sc i}, \h2, and C~{\sc ii$^*$} should be found in systems with 
21 cm absorption. The absence of C~{\sc i} and \h2 in the few systems with 
21 cm absorption is inconsistent with this prediction, again suggesting a local source of star light. 

\subsection{Effects of micro-turbulence and cosmic rays}

\begin{figure}
\centerline{\epsfig{file=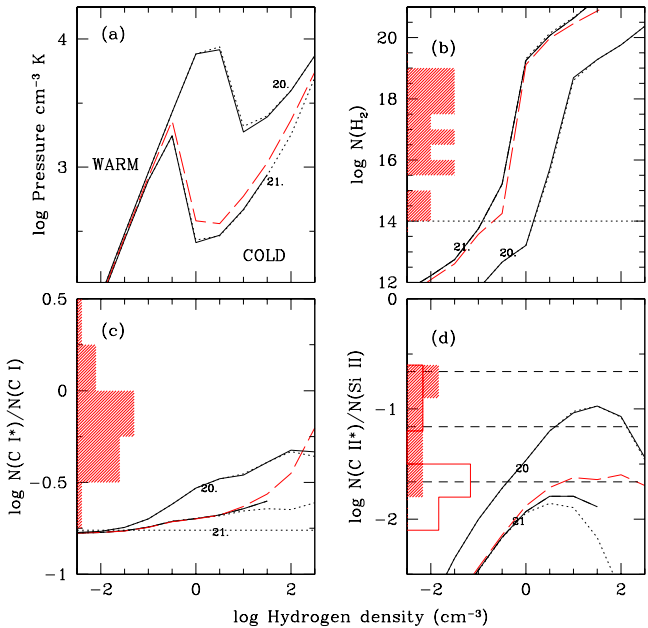,width=9cm,height=9cm}}
\caption {The effects of micro-turbulence 
\& cosmic ray ionization. The solid curves give the result without 
turbulence and  cosmic ray ionization when Z = 0.1 Z$_\odot$ and 
$\kappa = 0.01$. The short-dashed curves show the effects of  
3 \kms~turbulence. The long-dashed curves shows the 
effect of cosmic ray ionization (with a cosmic ray ionization rate of hydrogen 2.5 $\times10^{-17}$ s$^{-1}$).
The results are presented for two values of \N(H~{\sc i})(10$^{20}$, 10$^{21}$ cm$^{-2}$).
\label{fig4}}
\end{figure}

We have neglected
cosmic rays and turbulent motions in the models considered until now. 
In this
section we investigate whether the inclusion of these
processes help bring the Bgr models into agreement with observations.

The presence of micro-turbulence
increases the mean free path for
line photons and reduces line centre optical depth.
We consider a turbulent 
velocity corresponding to a Doppler {\it b} parameter of 3 
\kms. This shows the greatest possible effect since the typical measured {\it b} parameters of 
\h2 components are usually $\le 3$ \kms
%
(Srianand et al. 2005; Ledoux et al. 2003;
Petitjean et al. 2002; Srianand et al. 2000). 

We consider two column densities,
\N(H~{\sc i}) = 10$^{21}$ cm$^{-2}$ and 
\N(H~{\sc i}) = 10$^{20}$ cm$^{-2}$, 
along with $\kappa$ = 0.01, and {\it Z} = 0.1 {\it Z}$_\odot$.
Some results are presented in Fig.~\ref{fig4}.
\N(\h2) changes very little for \N(H~{\sc i}) = 10$^{21}$ cm$^{-2}$.
However, the molecular column density is slightly lower in the case 
of \N(H~{\sc i}) = 10$^{20}$ cm$^{-2}$.
We would need {\it b} $\gg$ 3 \kms~to produce a significant 
effect on \N(\h2).
%

[C~{\sc i}] 610$\mu$ is optically thick 
at high column densities and the fine-structure level populations 
are influenced by line trapping. Turbulence
increases the photon mean free path and reduces
this trapping.
%
%
The effect is clearly seen for the ratio \N(C~{\sc i$^*$})/\N(C~{\sc i}) when \N(C~{\sc i}$^*$)$\ge 3\times 10^{12}$ cm$^{-2}$, 
which occurs when \nh =10 cm$^{-3}$ (Panel (c) of Fig.~\ref{fig4}) for \N(H~{\sc i}) = 10$^{21}$cm$^{-2}$.
The effect is not seen significantly in the case of \N(H~{\sc i}) = 10$^{20}$
cm$^{-2}$ since 
large column densities of \N(C~{\sc i$^*$}) only occur for densities above
100 cm$^{-3}$. 
%
%
As pointed out before, our calculations produce higher \N(C~{\sc i}) than observed.
\N(C~{\sc i}) is invariably not saturated in most DLAs and $\lambda$610$\mu$ 
line trapping should not control the
\N(C~{\sc i$^*$})/\N(C~{\sc i}) ratio. Thus the inclusion of
micro-turbulence can not rectify the problems of the Bgr models
in simultaneously reproducing the \h2 and C~{\sc i} observations.

\par
The column density of C~{\sc ii}$^*$ is also affected by
the presence of turbulent motions for \N(C~{\sc ii$^*$}) $\ge 3\times 10^{13}$ cm$^{-2}$ due to optical depth in the [C~{\sc ii}]
158$\mu$ line.
The observed \N(C~{\sc ii$^*$}) is always higher than 3$\times10^{13}$ cm$^{-2}$
(Table 1 of Wolfe et al. 2003a; Srianand et al. 2005) in DLAs with \N(H~{\sc i})$\ge 10^{21}$ cm$^{-2}$.
Thus, line trapping effects may be important in producing the observed
excitation of the C~{\sc ii} fine-structure level.

%

Cosmic rays  
add heat to a highly ionized gas and produce secondary 
ionizations in a neutral gas. 
H$^+$ produced by cosmic ray ionization
can cause ortho-para conversion and thermalize this ratio
(Flower et al. 1994).
We consider a cosmic ray  
ionization rate equal to the Galactic background ionization 
rate ($\sim2.5\times10^{-17}~{\rm s}^{-1}$;  
Williams et al. 1998). These results are presented with
long-dashed lines in Fig~\ref{fig4}. 
The enhancement in the gas temperature increases the
column densities of C~{\sc i$^*$} and C~{\sc ii$^*$} in the
high-density gas. The pressure of the neutral gas increases due to cosmic ray heating as expected.
We also see a decrease in N(\h2) at a
given \nh~mainly  
because the increase in the gas temperature 
reduces the efficiencies in forming \h2.


Background cosmic ray ionization 
does not produce drastic changes for
the low-density gas 
where the Bgr dominates (Fig.~\ref{fig4}).  
A much larger cosmic ray ionization rate is needed to 
have an desired effect. 
%

\subsection{Summary:}

The main results for a cloud
irradiated by the meta-galactic UV background radiation are:
\begin{itemize}
\item{} The presence of the QSO dominated meta-galactic radiation field can 
maintain a \h2 abundance lower than the detection 
threshold for \nh$\le0.1$ cm$^{-3}$, irrespective of the dust content and \N(H~{\sc i}). The presence of any extra 
radiation field in addition to the meta-galactic radiation
field, or a slower \h2 grain formation rate, increases   
this critical density. Thus the absence of \h2 in 85 per cent of
DLAs is consistent with the low density models.

%

\begin{figure*}
\centerline{\epsfig{file=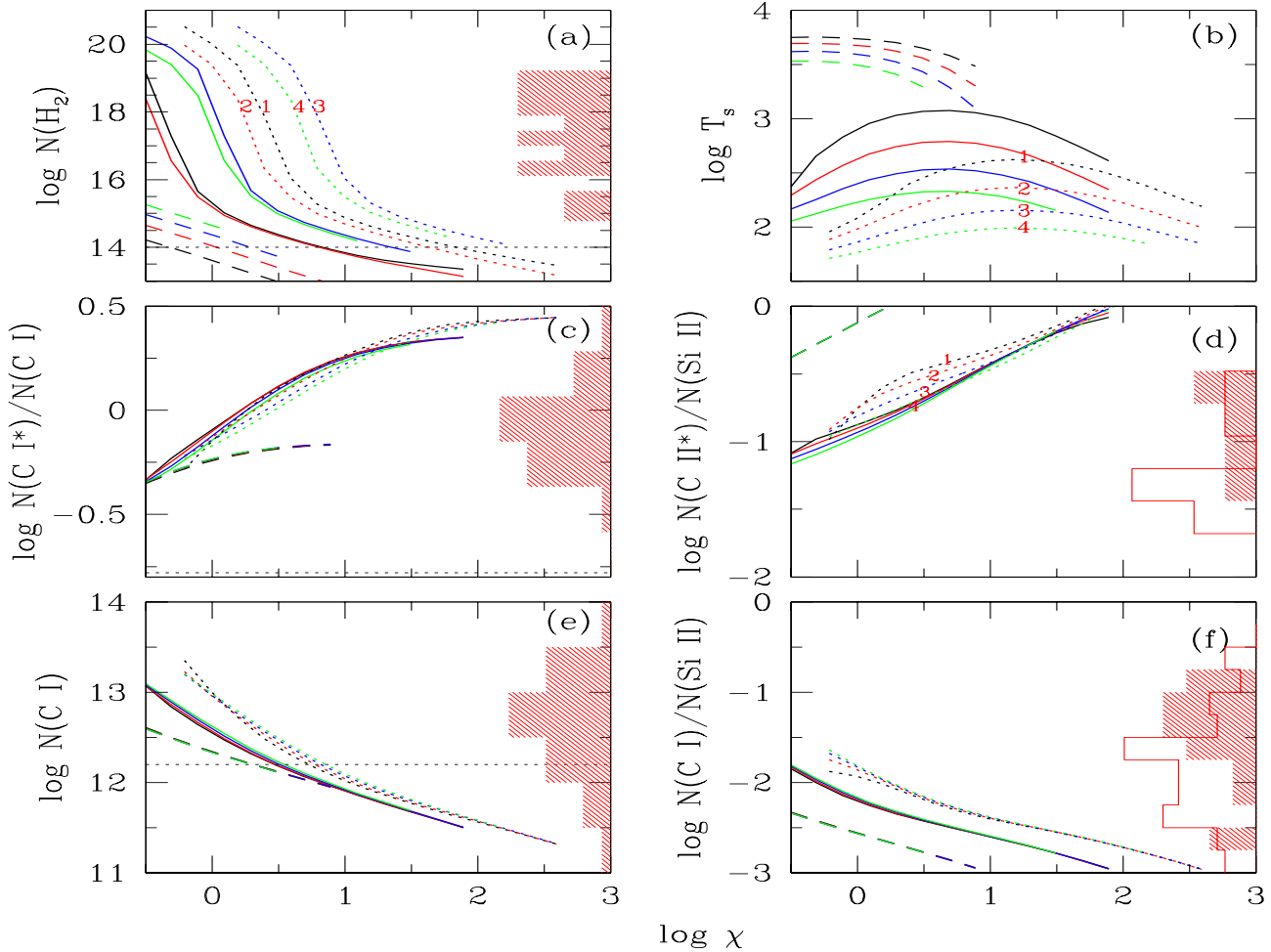,width=18cm,height=14cm}}
\caption { {\bf The effect of density:}
The results of calculations of a cloud in the
radiation field of a star with a surface temperature 40,000 K
(stellar case).
The cloud has log \N(H~{\sc i}) = 20.7, {\it Z} = 0.1 {\it Z}$_\odot$.
The long-dashed, continuous and short-dashed curves are for log \nh = 0.0, 1.0, and
1.7 respectively. The labels 1,2,3, and 4 in the short-dashed curves are
for log($\kappa$) = $-1.4,~-1.6,~-1.8$, and $-2.0$ respectively. 
We mark these  numbers only for the short-dashed curves. The dependences
on $\kappa$ has the same sense for other values of \nh~as
well. The results are presented as a function of $\chi$ (see Eq.\ref{eqchi}).
In each panel, the observed distributions are given as histograms
(see caption of Fig.~\ref{fig1} for details).
\label{figbb1}}
\end{figure*}
\item{} The detection of C~{\sc i} absorption
is inevitable whenever our line-of-sight passes through the CNM.
However, the density range that produces the observed 
\N(C~{\sc i}$^*$)/\N(C~{\sc i}) ratio also produces \N(C~{\sc i}) and
\N(\h2) higher than the observed values. An additional
source of radiation with energies $\le 13.6$ eV is needed to account 
for the low values of \N(\h2) and
\N(C~{\sc i}) in these systems.
\item{} Like \h2, the absence of 21 cm absorption in most of the high-{\it z} DLAs can be
naturally explained if DLAs originate mostly in the WNM.
A low-density gas, corresponding to a WNM, has a very large spin temperate  
({\it T}$_s$ $\ge$ 7000 K). If DLAs are dominated by
such a gas then 21 cm absorption will not be
detectable.
\item{}  A high-density gas has {\it T}$_s$ $\le$ 100 K.
This produces strong 21 cm absorption along with 
high values of \N(\h2) and \N(C~{\sc i}).
The fact that \N(\h2) and \N(C~{\sc i}) are not seen in the few
systems that do show 21 cm absorption suggests that an additional 
radiation field is present in these systems as well.
 This is consistent with results of detail investigations of 
individual systems available in the literature
(see references given in Section 3.3.1).

\item{} The too-large \N(\h2) and \N(C~{\sc i}) column densities 
predicted at higher densities cannot
be explained by micro-turbulent motions (up to 3 \kms) or cosmic
ray heating. However, the
non-thermal motions affect the 
fine-structure level populations when the infrared lines become
optically thick. The observed log \N(C~{\sc ii$^*$}) is $\ge 13.5$
for \N(H~{\sc i})$\ge 10^{21}$ cm$^{-2}$. The effects of line trapping
of [C~{\sc ii}] $\lambda158$ becomes very important in these systems 
whenever the turbulent motions are small.
\end{itemize}
In  the following section we explore the possibility of using
{\it in situ} star formation to prevent the formation of too-large 
\N(\h2) in high-density systems. We wish to point-out that inclusion of 
radiation from the Lyman Break Galaxies (LBGs) can increase the
flux of Lyman Warner band photons in the UV background by upto a 
factor 10 (see Haardt \& Madau (2001)). This will make the
Bgr radiation roughly 5 times less than the Galactic mean field.
However, Section 4 shows that the required radiation field is 
much higher than this enhanced Bgr.
In addition, this enhanced UV field will also produce 
bimodel distribution of N(\h2) contrary to what has been 
observed.

\section {Ionization by young stars:}

\begin{figure*}
\centerline{\epsfig{file=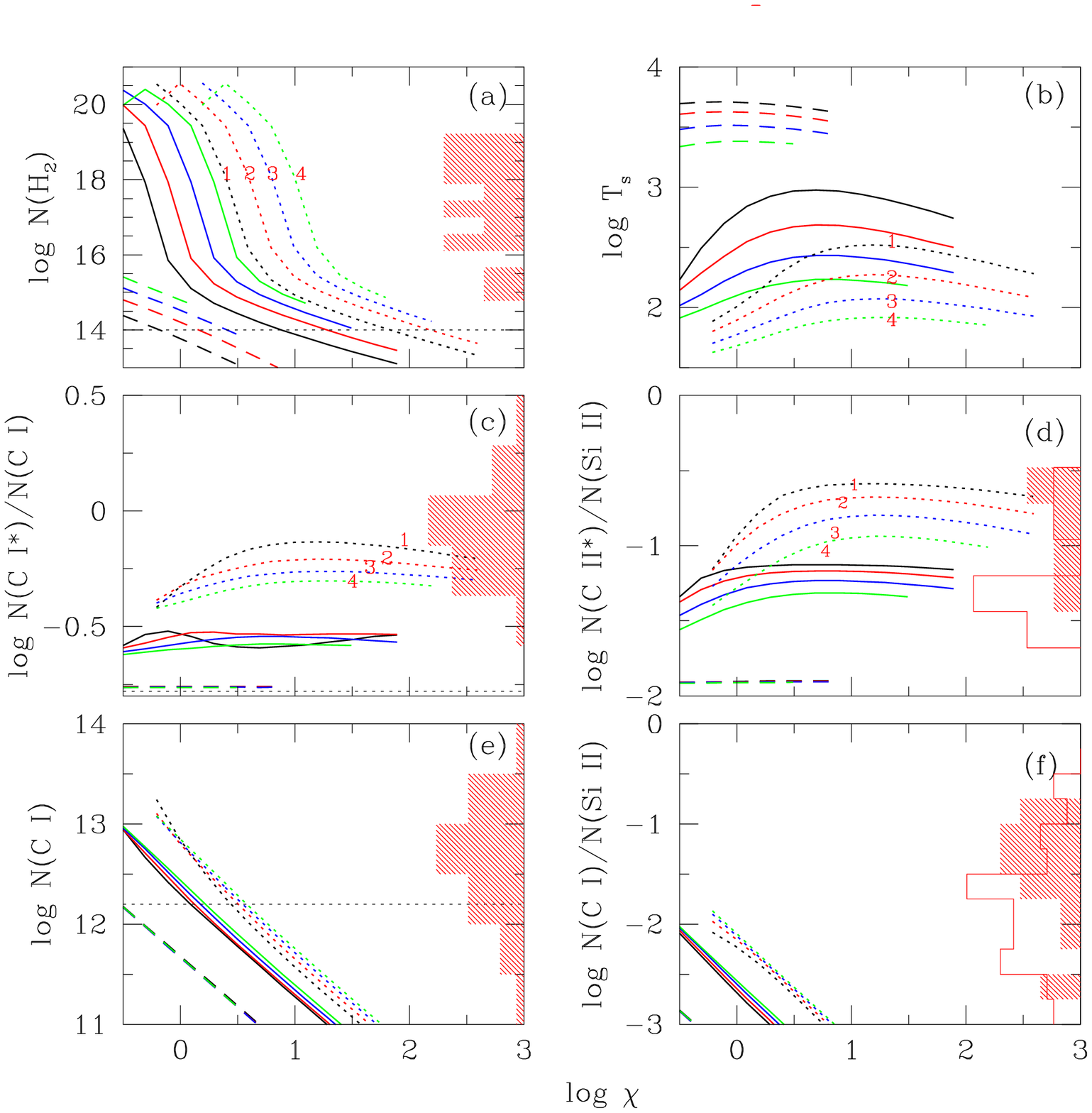,width=18cm,height=14cm}}
\caption {The results of calculations of a cloud in the
radiation field of a star with a surface temperature 40,000 K and 
attenuated by \N(H~{\sc i}) = 20 cm$^{-2}$ (diffuse case). Rest are same as in
Fig.~\ref{figbb1}
\label{figbb1e}}
\end{figure*}
\begin{figure*}
\centerline{\epsfig{file=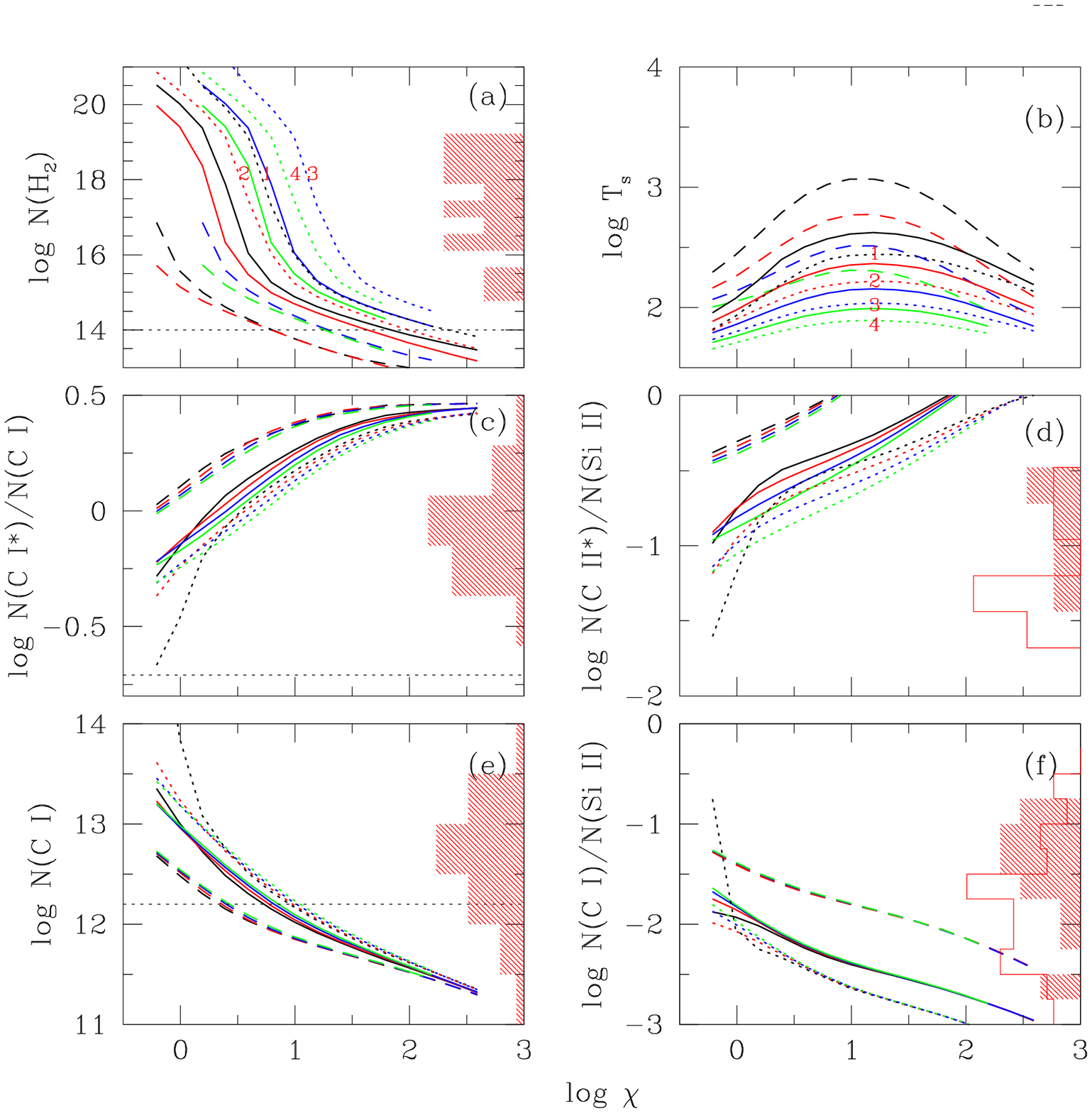,width=18cm,height=14cm}}
\caption { {\bf Effects of column density:}
The results of calculations of a cloud in the
radiation field of a star with surface temperature 40,000 K (stellar case).
The cloud has \nh=50 cm$^{-3}$, {\it Z} = 0.1 {\it Z}$_\odot$.
The long-dashed, continuous and short-dashed  curves are 
for log \N(H~{\sc i}) = 20.0, 20.7 and
21.0 respectively. The labels 1,2,3, and 4 in the short-dashed curves are
for log($\kappa$) = $-1.4,~-1.6,~-1.8,$ and $-2.0$ respectively. 
We mark these  numbers only for the short-dashed curves. The dependences
on $\kappa$ has the same sense for other values of \N(H~{\sc i}) as
well. The histograms are as explained in Fig.~\ref{fig1}.
\label{figbb1a}}
\end{figure*}
\begin{figure*}
\centerline{\epsfig{file=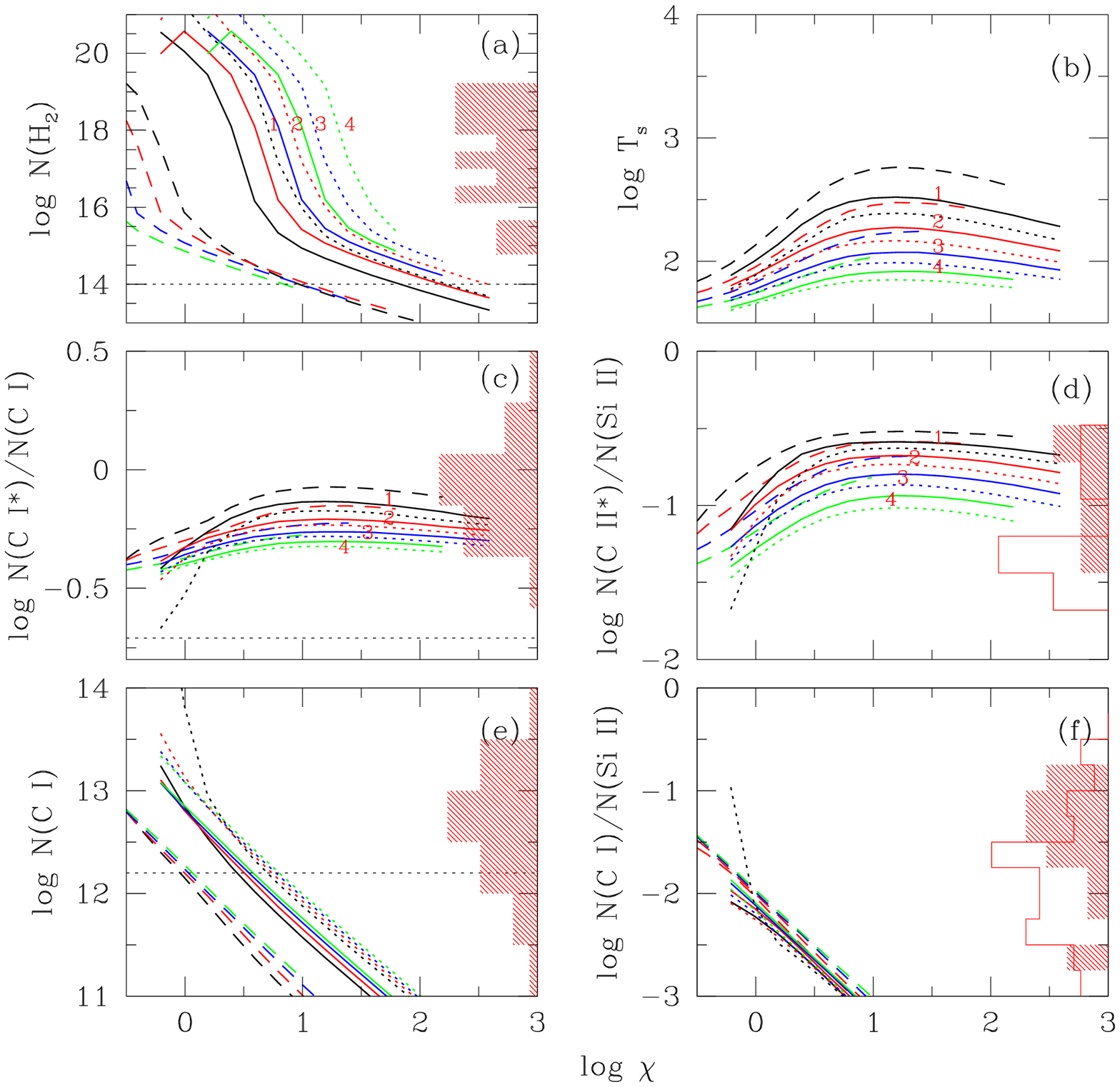,width=18cm,height=14cm}}
\caption { The results of calculations of a cloud in the
radiation field of a star with surface temperature 40,000 K
attenuated by \N(H~{\sc i}) = 20 cm$^{-2}$, our diffuse case. Rest are same as in
Fig.~\ref{figbb1a}
\label{figbb1ae}}
\end{figure*}


%
We add a stellar radiation field using a 40,000 K 
Kurucz' model atmosphere in addition to the metagalactic radiation field 
discussed above. We consider direct stellar radiation and stellar 
continuum attenuated by N(H~{\sc i}) = $10^{20}$ cm$^{-2}$.
These two cases are denoted by ``stellar" and ``diffuse", 
as discussed in Section 2.5.
We showed that the 
meta-galactic UV background is sufficient to suppress
the formation of \h2 in low-density gas (i.e \nh$\le0.1~{\rm cm}^{-3}$). 
Thus, in this section
we mainly concentrate on the high-density gas needed to account for the observed \N(C~{\sc i$^*$})/\N(C~{\sc i}) ratio.
We consider clouds with three values 
of \N(H~{\sc i}) ($10^{20}$, $10^{20.7}$, and 10$^{21}$ cm$^{-2}$),
three values of \nh (1, 10, and 50
cm$^{-3}$), and four values of $\kappa$ ($10^{-2},$ $10^{-1.8},
$ $10^{-1.6}$, and $10^{-1.4}$). In all of these calculations we assume
{\it Z} = 0.1 {\it Z}$_\odot$, turbulent velocity {\it b} = 3 \kms.
Cosmic-ray heating is not considered in the models described below.
%
\par
%
We present the results of model calculations as a function of
$\chi$ (as defined in Eq.~\ref{eqchi}).
The calculations for stellar case will
have more high-energy photons than the diffuse one for the same value of $\chi$. 
Therefore, for a given \N(H~{\sc i}),
the gas will be hotter and more ionized at the
illuminated side of the cloud for the stellar case 
(see Fig.~\ref{depth}).
We present the results of our calculations 
for the stellar and diffuse case for
\N(H~{\sc i}) = 5 $\times10^{20}$ cm$^{-2}$ in 
Figs.~\ref{figbb1} and \ref{figbb1e} respectively.
The effects of changing  \N(H~{\sc i}) are 
shown in Figs.~\ref{figbb1a} and \ref{figbb1ae}. 
%

\par

\subsection{\h2 abundance:}
\begin{table*}
{\tiny
\caption{Results for clouds irradiated by starlight. A-E are the stellar case and AE-EE are the diffuse case}
\begin{tabular}{lcccccccccc}
\hline
&\multicolumn {10}{c}{Models} \\
Parameters   & A & B & C &D&E& AE & BE &CE&DE&EE\\
\hline
log N(H~{\sc i})     & 20.7   & 20.7 & 20.7 &20.0&21.0& 20.7  & 20.7  &20.7&20.0&21.0\\ 
\nh(cm$^{-3}$)       & 1      & 10   & 50   &  50& 50 & 1     &  10   &50. &50  & 50 \\
log N(H~{\sc i}(ext))& ....&.... & .... &.... &....  &20.0 &20.0& 20.0  & 20.0  &20.0\\
$\chi $              & $\le 3$& 1,30 & 3,100&0.5,35&4,400&$\le10$& 1,100 &3,200&0.3,22&10,475\\
T$_s$ (K)            & 3150,5600& 100,1000 & 56,316&104,1260&48,275 &2290,5248&90,900&54,339&40,560&66,195\\ 
$\tau_v({\rm 21~cm})/f$&0.09,0.05 &2.74,0.27&4.90,0.87&0.53,0.04&11.42,2.00& 0.12,0.05& 3.04,0.30& 5.07,0.81&1.37,0.10 &0.83,0.28\\
log N(C~{\sc i})/N(Si~{\sc ii}) &$-2.6,-2.0$ &$-2.6,-2.0$&$-2.5,-1.8$&$-1.8,-1.0$&$-3.0,-2.5$&$-3.8,-2.5$&$-4.0,-2.04$&$-3.7,-2.0$&$-2.5,-1.0$&$-3.7,-2.0$\\
log {N(C~{\sc i}$^*$)/ N(C~{\sc i})}&
 $-0.76$,0.0 & $-0.3$,0.3 &$-0.2$,0.4 &$-0.1,0.44$&$0.0,0.40$&$\sim-0.76$&$-0.6,-0.5$&$-0.4,-0.1$&$-0.4,-0.1$&$-0.26,-0.44$\\
log N(C~{\sc ii$^*$})/N(Si~{\sc ii}) & $-1.8,-1.3$ & $-0.5,-1.0$& $-0.2,-0.8$ & $-0.5,0.0$&$-0.9,-0.1$&$\sim-1.9$&$-1.3,-1.1$ & $-0.9,-0.6$ &$-1.5,-0.5$&$-1.08,-0.66$\\
%
log N(C~{\sc i}(tot)) &12.0,13.4 &11.9,13.0 & 11.8,12.8& 11.0,13.4&11.8,13.0&11.0,13.0&11.0,13.0&11.0,13.0&11.2,13.0&10.6,12.0\\
log N(O~{\sc i$^*$})   &11.2,11.3 & 11.3,11.9 & 11.5,12.2 &11.0,11.3 & 10.5,11.7&11.0,11.2&10.0,11.5&10.0,12.0&10.0,11.5&10.5,12.0\\
log N(O~{\sc i$^{**}$})&11.2,11,3 & 11.3,11.9 & 11.6,12.2 &11.0,11.3 & 10.5,11.5&11.0,11.3&11.0,11.53&110.0,11.5&0.0,11.3&10.0,11.7\\
log N(Si~{\sc ii$^*$}) & $\le10.2$& 10.5,11.0 & 10.8,11.7 &$\sim9.5$ & 9.6,10.2 &9.0,10.3& 9.9,10.0&9.0,10.5&9.0,10.6&9.0,10.5\\
%
\hline
\end{tabular}
\label{table1}
}
\end{table*}
%
%
Panel (a) of Fig.~\ref{figbb1} shows the predicted column density of \h2 
as a function of $\chi$ 
for 
the stellar case, log~\N(H~{\sc i})=20.7, and 3 values of
\nh (long-dashed, continuous and short-dashed curves are for
\nh = 1, 10 and 50 cm$^{-3}$ respectively). For each values of
\nh the results are presented for 4 different $\kappa$.
For a given \nh~and $\chi$, a higher dust content $\kappa$ produces  
a lower \N(\h2). Naively we would expect \N(\h2) to increase with increasing $\kappa$ due to an enhanced probability of H 
striking a grain, and increased shielding. 
However, the gas temperature increases for higher  
$\kappa$ due to additional grain photo-electric
heating (see the next section). This reduces the \h2 formation rate 
as shown in Fig.~1 of Cazaux and Tielens (2002),
so, \N(\h2) decreases.
\par
The calculations reproduce the observed range of \N(\h2) for 1$\le\chi\le$100,
\nh~ $\sim$ 10$-$50 cm$^{-3}$, and log~\N(H~{\sc i}) = 20.7 for the stellar case.
The range becomes 1$\le\chi\le$300 for the 
diffuse case (Panel (a) of Fig.~\ref{figbb1e}).
We find that with \nh=1 cm$^{-3}$, \h2 should be detectable when 
$\chi\le 3$ and $\chi\le 10$ for the stellar 
and diffuse case respectively. Clearly, for a moderate local radiation field, 
$\chi\simeq 1-10$, \h2 will be detectable for \nh$\ge$1 cm$^{-3}$ 
and \N(H~{\sc i}) $\ge5\times10^{20}$ cm$^{-2}$.  
Higher \nh~ is needed to produce detectable \N(\h2) for a lower \N(H~{\sc i})
(Panels (a) in Figs.~\ref{figbb1a}
and \ref{figbb1ae}). The range of $\chi$ that is consistent with the 
observed range of \h2 is summarised in Table~\ref{table1} for all 
the scenarios discussed in this work.  
Observations of atomic fine-structure lines will further narrow down this range.

\par
\subsection{Spin temperature and 21 cm optical depth:}
Panels (b) of Figs.~\ref{figbb1} and \ref{figbb1e}
show the predicted {\it T}$_s$ for log~N(H~{\sc i}) = 20.7, 3 values
of \nh (Long-dashed, continuous and short-dashed curves are for
\nh=1,10 and 50 cm$^{-3}$ respectively) and 4 values of $\kappa$.  
It is clear from all these panels that
for a given \nh (curves with similar line style), $T_s$ increases with 
increasing $\kappa$ mainly due to photoelectric 
heating by dust grains (labels 1, 2, 3 and 4 on the short-dashed
curve show models with $\kappa$ in the increasing order).
We also notice that for  a given $\kappa$ (say top-most
curve with a given line-style) and $\chi$ the models with
higher \nh have lower {\it T}$_s$. This effect is very prominent
in the stellar case. This is mainly because {\it T}$_s$ is very
large at the illuminated side of the gas in the stellar case. 
Panels (b) of Figs.~\ref{figbb1a} and \ref{figbb1ae} show 
the results with \nh = 50 cm$^{-3}$ for
3 different values of N(H~{\sc i}) (long-dashed, continuous and short-dashed 
curves are respectively for log~{N(H~{\sc i})}= 20, 20.7 and 21)
for the range of $\kappa$. It is clear 
that for a given $\kappa$ (say top most curves for different
line-styles), clouds with lower \N(H~{\sc i}) will 
have higher $T_s$ and hence lower $\tau_v({\rm 21~cm})$.
In the diffuse case
(from Figs.~\ref{figbb1e} and \ref{figbb1ae}) we notice that for a given 
\nh and $\kappa$, {\it T}$_s$ gradually increases with $\chi$ and saturates
to a constant for large values of $\chi$. The increase in $T_s$ with lower
value of $\chi$ is the effect of photo-electric heating. At larger $\chi$
the grains become highly charged and total heating rate will become independent
of $\chi$ (Bakes \& Tielens, 1994; Weingartner \& Draine (2001a)). Thus
$T_s$ becomes independent of $\chi$ at large $\chi$.

\par
The range of predicted $T_s$  
in the range of ${\rm \chi}$ constrained by the \h2 observations
is summarised in Table.~\ref{table1}. This table also gives the
expected 21 cm optical depth obtained
using Eq.~\ref{eqn1}. {It is clear from the table that
for a high-density gas 
(i.e \nh$\ge$10 cm$^{-3}$) and log~\N(H~{\sc i})$\ge20.7$,
21 cm absorption should be detectable
with an optical depth of $\tau_v({\rm 21~cm})/f\ge0.27$. 
Especially for high $\kappa$,
the 21 cm optical depth becomes as low as 0.05, even when
\nh=50 cm$^{-3}$, for log~\N(H~{\sc i}) = 20. 
{Thus, the absence of 21 cm absorption in  
high-density (or low temperature) systems will either mean that 
the average N(H~{\sc i}) along the radio source is much less
than N(H~{\sc i}) seen along the optical sight-line (Wolfe et al. 2003a)
or that the actual \N(H~{\sc i}) in the high-density component is lower
(Kanekar \& Chengalur, 2003).
Absence of \h2 in the systems that show 21 cm absorption with 
low $T_s$ will indicate a radiation field much higher (i.e $\chi>>1$).
}
Our calculations also suggest that it is possible to
detect \h2 with low $\tau_v({\rm 21 cm})$ for moderate \nh
(see Models A \& AE in Table.~\ref{table1}). For example, when 
log \N(H~{\sc i}) $=20.7$ and \nh $\simeq 1$ cm$^{-3}$, the expected spin 
temperature is high (2300-5600 K) and $0.05\le\tau_v({\rm 21 cm})/f\le 0.12$. 
Thus, 21 cm absorption will either be weak or undetected in cases 
where \h2 is detectable. Thus, the fine-structure 
excitation of C~{\sc i} or C~{\sc ii} with  systems with 
detectable 21 cm absorption  
will lead to a better understanding of physical conditions 
in the gas. This is detailed in the following sections.}
%

%
\subsection{C~{\sc i} absorption: detectability and level of ionization:}
\begin{figure*}
\centerline{\epsfig{file=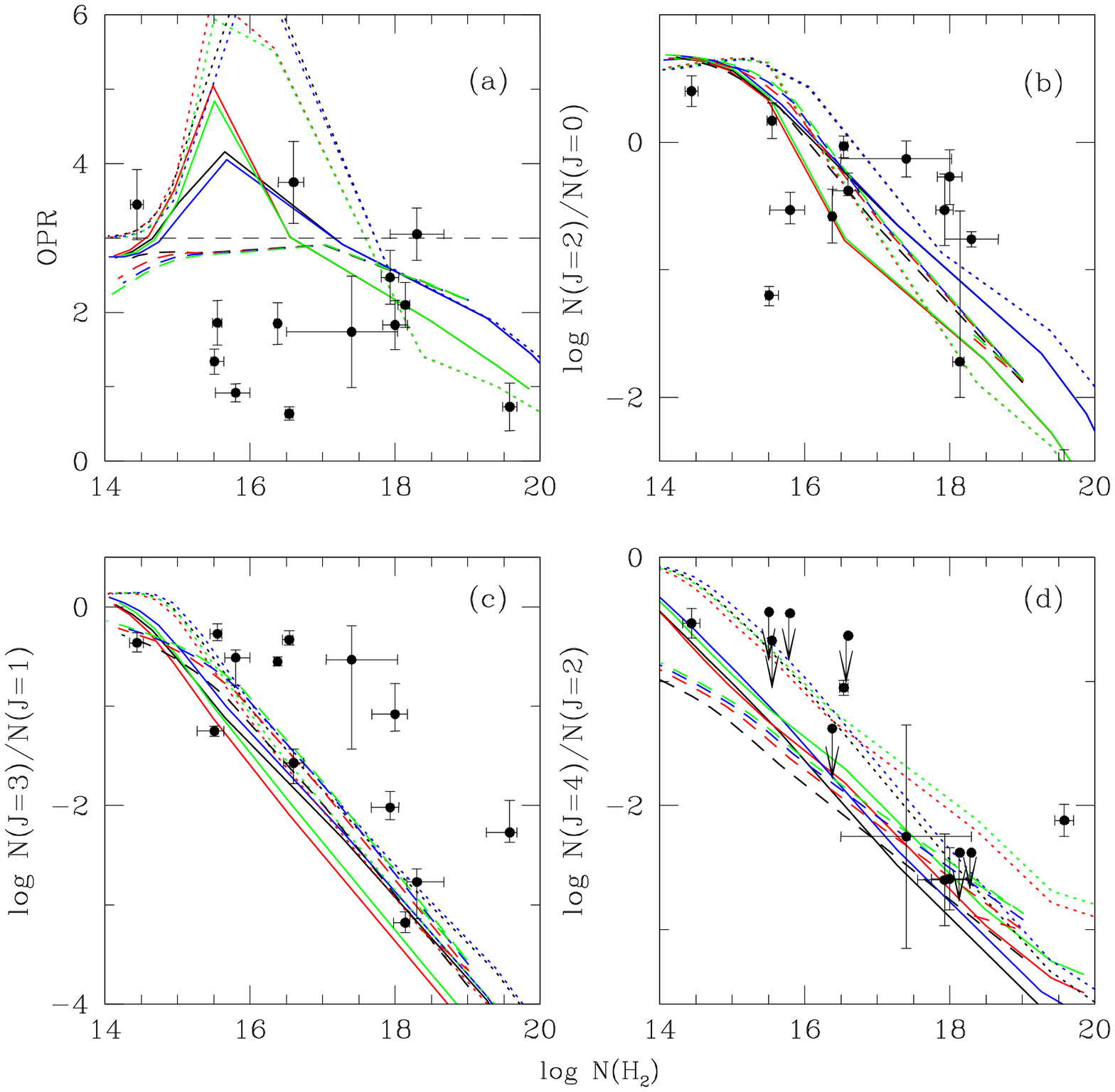,width=18cm,height=15cm}}
\caption {The ratio of column densities of \h2 in different
rotational levels are shown as a function of total \h2 column density for
the stellar case.
The points in the figures give the observed data (Ledoux
et al. 2003). 
In all these calculations we assume the metallicity to be 0.1 Z$_\odot$,
log~N(H~{\sc i})=20.7, and log $\kappa$ is varied between
$-2.0$ and $-1.4$ (different curves with same line-style). 
The long-dashed, continuous and short-dashed curves are
for \nh = 1, 10 and 50 cm$^{-3}$ respectively.
\label{figbb2}}
\end{figure*}

\begin{figure*}
\centerline{\epsfig{file=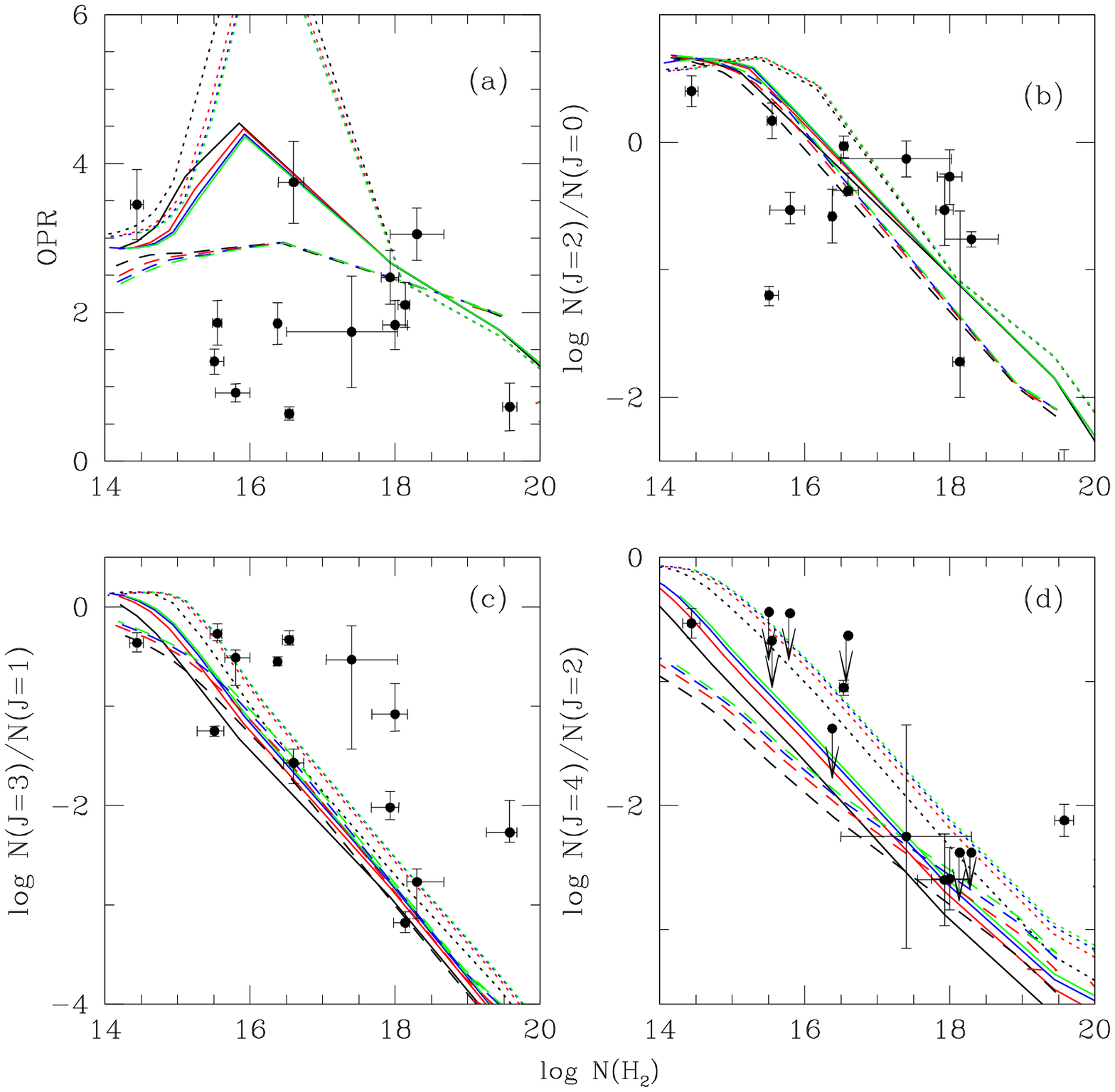,width=18cm,height=15cm}}
\caption {
Same as Fig.~\ref{figbb2} but diffuse ionizing radiation is used
in these calculations.
\label{figbb2e}}
\end{figure*}

%


{ Panels (e) and (f) of Figs.~\ref{figbb1} and~\ref{figbb1e} plot 
\N(C~{\sc i}) and \N(C~{\sc i})/\N(Si~{\sc ii}) as
a function of $\chi$ for stellar and diffuse cases respectively with
log~\N(H~{\sc i})=20.7. It is clear from the figure that the
predictions for different $\kappa$ (curves with same line-style) 
are very much identical.} 
Both \N(C~{\sc i}) and  \N(C~{\sc i})/\N(Si~{\sc ii})  
are lower than those produced by the 
Bgr (Fig.~\ref{fig1}) due to the presence of additional
ionizing photons. 
%
A minor  reduction in \N(C~{\sc i}) and \N(C~{\sc i})/\N(Si~{\sc ii}) for a given 
\nh~and $\kappa$ is noted for the diffuse case compared
to the stellar case. This
can be easily
understood using Fig.~\ref{depth}. N(Si~{\sc ii}) is
nearly identical for both the radiation fields. Whereas N(C~{\sc i})
is less in the diffuse case as $n_e$ and $T$ are lower
in these models compared to stellar case (see panel (b) of Fig.~\ref{depth}).
%
%

\par
Here we concentrate on systems with \h2 detections. The predicted 
\N(C~{\sc i}) is well below the detection limit for log~\N(H~{\sc i}) = 20.7 
and $\chi\ge10$ for the range of \nh~and the two stellar continua considered here.
Thus, these models can explain the weak or non-detection of
C~{\sc i} absorption in some of DLAs that show strong \h2 absorption
(\zabs = 3.025 toward Q~0347$-$383 with log~\N(H~{\sc i})= 20.56;
\zabs = 2.595 toward Q~0405$-$443 with log~\N(H~{\sc i})= 20.90; and 
\zabs = 2.811 toward Q~0528$-$250 with log~\N(H~{\sc i})= 21.10).
\N(C~{\sc i})/\N(Si~{\sc ii}) is lower than the measured ratio
(in DLAs that show both \h2 and C~{\sc i} absorption) for the 
range in $\chi$ allowed by \h2 for both the diffuse and stellar 
case, and log~\N(H~{\sc i}) = 20.7. 
Based on the trend seen in Figs.~~\ref{figbb1} 
and~\ref{figbb1e} we may need \nh$\ge$ 50 cm$^{-3}$  and $\chi\le10$
to explain the observed range in \N(C~{\sc i})/\N(Si~{\sc ii})
for \h2 components that show detectable C~{\sc i} absorption.
However, such a model will over produce \N(\h2). This inconsistency
can be solved by using a lower value of \N(H~{\sc i}) and a higher \nh
(see Panel (f) in Figs.~\ref{figbb1a} and \ref{figbb1ae}).
{In these plots long-dashed, continuous and short-dashed curves 
are the results for \nh=50 cm$^{-3}$ with log~N(H~{\sc i}) = 20.0, 20.7, and
21.0 respectively. 
}
The total \N(H~{\sc i}) measured for \zabs = 1.968 toward
Q~0013$-$004 and \zabs = 2.087 toward 1444+014 are consistent with
log~\N(H~{\sc i})$\le$20 in the \h2 (and C~{\sc i}) components.
The \zabs = 1.962 system toward Q~0551-366 that shows three
\h2 components has a total log~\N(H~{\sc i})=20.5. The \zabs=1.973
system toward Q~0013-004 has 15 well detached C~{\sc i}
components with a total \N(H~{\sc i})=20.8. Clearly a low
N(H~{\sc i}) is probable in the components that show \h2 and 
C~{\sc i} absorption.  
%
%
%
%
%
A cloud with
a moderate radiation field (i.e $\chi\le10$), log~\N(H~{\sc i})=20.0, and \nh = 50 cm$^{-3}$ reproduces the observed range
of \N(C~{\sc i})/\N(Si~{\sc ii}), \N(\h2) and \N(C~{\sc i}). 
%
%
%

\subsection{C~{\sc i} fine-structure excitation:}
In this section we consider the fine-structure excitation of C~{\sc i}.
Panel (c) of Figs.~\ref{figbb1} and~\ref{figbb1e} shows 
\N(C~{\sc i$^*$})/\N(C~{\sc i}) as a function of $\chi$
for log~\N(H~{\sc i})=20.7. 
This ratio should be independent of $\chi$ if 
UV pumping is negligible because $T_s$ depends mainly on density and 
is roughly independent of $\chi$.
This happens for the diffuse case (Panel (c) of 
\ref{figbb1e}). {Also the effect of $\kappa$
is clearly evident in this case}.
However, in the stellar case, we find that the predicted 
\N(C~{\sc i$^*$})/\N(C~{\sc i}) increases with increasing $\chi$.  
We also notice that for a given \nh~and $\chi$, the stellar 
case (Panel c in Fig.~\ref{figbb1a}) with a lower 
\N(H~{\sc i}) produces a higher value of \N(C~{\sc i$^*$})/\N(C~{\sc i}).
However, the dependence on \N(H~{\sc i}) is very weak in clouds   
irradiated by the diffuse radiation field (see panel c in Fig.\ref{figbb1ae}).
{This implies that the ratio 
\N(C~{\sc i$^*$})/\N(C~{\sc i}) will depend only on density for the diffuse 
case. However,
the ratio
\N(C~{\sc i$^*$})/\N(C~{\sc i}) will depend on the strength
of the radiation field (also see panel (d) in  Fig.~\ref{depth})
in the stellar case.}
\par
Now we focus on systems that show detectable 
C~{\sc i} and \h2.  
The predicted value of \N(C~{\sc i}$^*$)/\N(C~{\sc i})
is more sensitive to \nh~and weakly depends on \N(H~{\sc i}) and
$\chi$ for the diffuse case. The observed \N(C~{\sc i}$^*$)/\N(C~{\sc i}) is consistent
with 10$\le$\nh$\le100$ cm$^{-3}$. In the 
stellar case \N(C~{\sc i}$^*$)/\N(C~{\sc i}) depends on
\nh, N(H~{\sc i}) and $\chi$. Clouds with log~\N(H~{\sc i}) = 20.7
reproduce the observed range in \N(C~{\sc i}$^*$)/\N(C~{\sc i}) for
1$\le$\nh$\le$50 cm$^{-3}$ and the values of $\chi$ constrained by the \h2
observations. As noted before, however, these models fail to reproduce
the observed \N(C~{\sc i})/\N(Si~{\sc ii}). 
%
%
%
%
%
Thus we require
low \N(H~{\sc i}) ($\simeq 10^{20}$ cm$^{-3}$), high \nh ($\simeq 50$ cm$^{-3}$), 
and 
low $\chi$ ($\le10$)
components in order to be consistent with the observed
 \N(C~{\sc i}) and \N(C~{\sc i})/\N(Si~{\sc ii}) ratio.

\subsection{C~{\sc ii} fine-structure excitation:}

Here we discuss the predicted C~{\sc ii$^*$} in 
detail. Panel (d) of Figs.~\ref{figbb1} and \ref{figbb1e} shows 
\N(C~{\sc ii$^*$})/\N~({Si~{\sc ii})  
for log~\N(H~{\sc i}) = 20.7 as a function of $\chi$. 
The shaded histogram  is the observed distribution
of the systems with \h2 components and the non-shaded histogram 
represents those systems that do not show detectable \h2 and 
C~{\sc i}. 

\par
{In the stellar case \N(C~{\sc ii$^*$})/\N~({Si~{\sc ii})
is higher for higher $\chi$ (see panel (d) in Fig.~\ref{figbb1}), 
and for a given $\chi$ 
the excitation is more for lower density (long-dashed, continuous
and short-dashed lines are for \nh = 1, 10, and 50 cm$^{-3}$ respectively).
The ratio depends only weakly on $\kappa$ (different curves with 
different line-style) in the stellar case.}}
{
All these trends are mainly because, for a fixed \N(H~{\sc i}),
a considerable fraction of C~{\sc ii} will originate from regions where hydrogen is ionized.
The fraction of C~{\sc ii} originating from a hot ionized
gas is higher in the case of lower \nh~and so
the ratio is higher.
This happens for the lower \N(H~{\sc i}) case also (see panel (d)
in Fig.~\ref{figbb1a} where long-dashed, continuous and short-dashed
curves are for log~N(H~{\sc i}) = 20., 20.7 and 21 respectively).
}

\par
{ Results for the diffuse case are summarised in panels (d) of
Figs. \ref{figbb1e},\ref{figbb1ae}. It is clear that 
for a given \nh, N(H~{\sc i}), and $\kappa$ the ratio
increases with increasing $\chi$ when the $\chi$ is small.
However, at larger $\chi$ the ratio becomes independent of
$\chi$. This is due to the grain heating 
saturation for highly charged grains at high $\chi$ (see
Weingartner \& Draine, 2001a). This is also the reason
for the lack of dependence of spin temperature on $\chi$
(see previous section). Thus, at high $\chi$ the ratios
mainly depend on \nh~in the diffuse case. The
models presented by Liszt (2002)
use the fitting function given by
Bakes \& Tielens (1994) and show a monotonic increase in
N(C~{\sc ii$^*$})/N(C~{\sc ii}) with an increase in $\chi$.
Weingartner \& Draine (2001a) (see figure 15 in their
paper) show that the fitting function given by Bakes \& Tielens (1994)
over produce the photo-electric heating  at high $\chi$ and
because of this there will be an increase in N(C~{\sc ii$^*$})/N(Si~{\sc ii})
with increasing $\chi$ even at large values of $\chi$. 
As our treatment is very close to that of Weingartner \& Draine (2001a)
we clearly see the effect of saturation of photo-electric heating by dust grains
in our models.
}
\par
First we will concentrate on the systems with \h2 detections.
In the diffuse case the range in \nh~that is
consistent with \N(C~{\sc i$^*$})/\N(C~{\sc i}) also reproduces 
the observed range in  \N(C~{\sc ii$^*$})/\N(S~{\sc ii}).
In the stellar case with log~\N(H~{\sc i}) = 20.7 
the observed distribution of \N(C~{\sc ii$^*$})/\N(Si~{\sc ii})  
is consistent with \nh~in the range of 10-50 cm$^{-3}$.  The models
with lower \N(H~{\sc i}) tend to produce higher \N(C~{\sc ii$^*$})/\N(Si~{\sc ii})
for a given \nh~and $\kappa$. A cloud with the low \N(H~{\sc i}) and
high \nh~that are required to reproduce N(C~{\sc i})/N(Si~{\sc ii})
will over produce \N(C~{\sc ii$^*$})/\N(Si~{\sc ii}).
Thus the observed \N(C~{\sc ii$^*$})/\N(Si~{\sc ii}) seems
to favor the diffuse radiation field. 
This produces a model that reproduces all the other observations.

Now we concentrate on systems without \h2 detections but
showing C~{\sc ii$^*$} absorption. {There are two possibilities
for the absence of \h2 in these systems: Either (i) the gas has lower density and so is 
partially ionized with a higher temperature or
(ii) the gas is at a high density in a strong UV field.} 
{ In the diffuse case the ratio N(C~{\sc ii$^*$})/N(Si~{\sc ii}) measured
in systems without \h2 are consistent with \nh in the range of 1-10 
cm$^{-3}$ (see panel (d) of Figs. \ref{figbb1e}). }
Srianand et al. (2005) pointed out that Al~{\sc iii}
absorption seen in these systems could be a useful indicator of the
ionization of the gas. We will return to this issue
while discussing the predicted \N(Al~{\sc iii})/\N(Al~{\sc ii}) ratio
(see Section 4.7).

\subsection{Rotational excitation of \h2:}

\begin{figure*}
\centerline{\epsfig{file=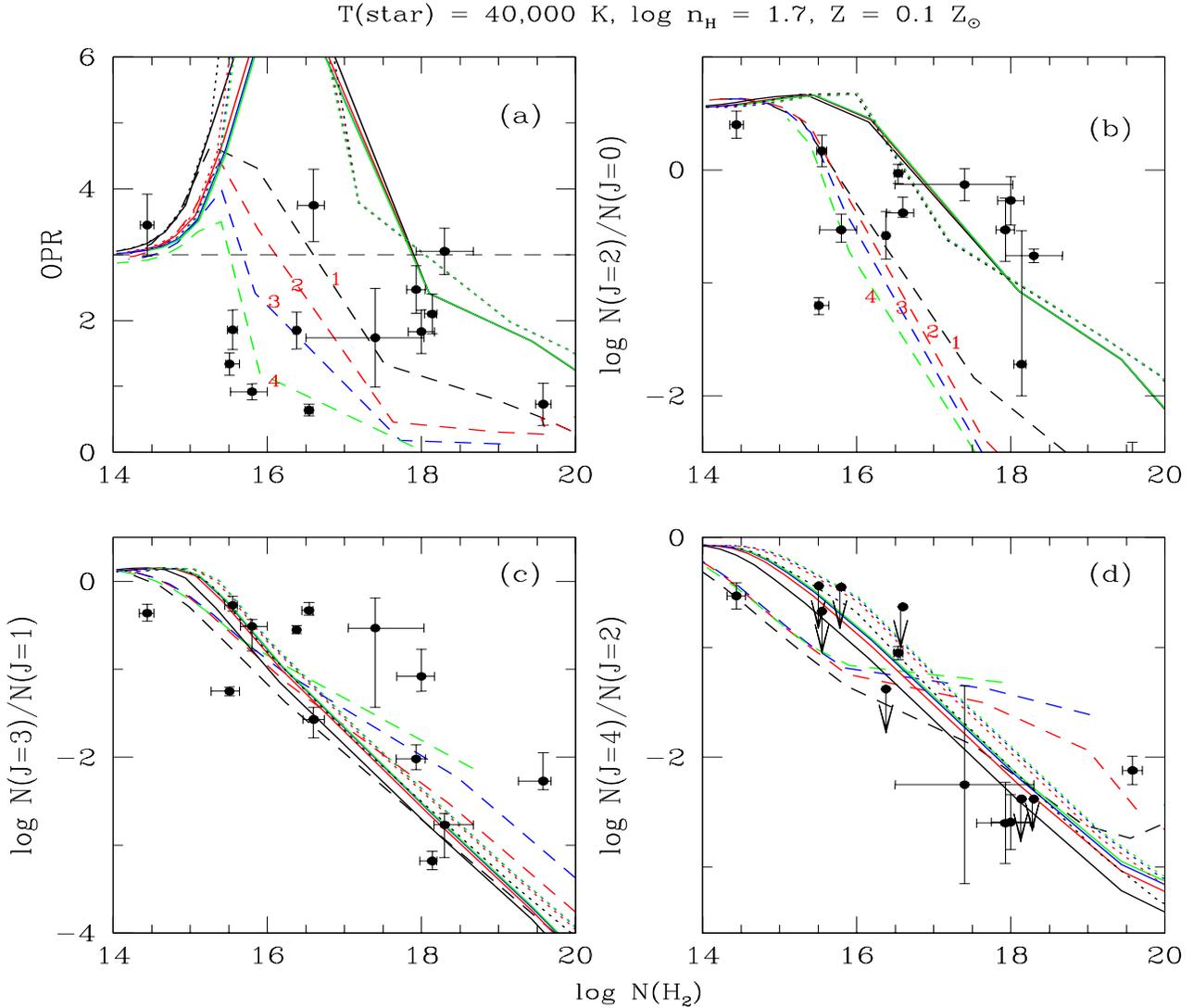,width=18cm,height=15cm}}
\caption {The ratio of column densities of \h2 in different
rotational levels are shown as a function of total \h2 column density
for diffuse case.
The points in the figures give the observed data (Ledoux
et al. 2003).  
In all these calculations, we assume the metallicity to be 0.1 {\it Z}$_\odot$,
 \nh= 50 cm$^{-3}$, and log $\kappa$ is varied between
$-2.0$ and $-1.4$. The labels 1, 2, 3 and 4 in long-dashed curves are 
for log($\kappa$) = $-1.4$, $-1.6$, $-1.8$ and $-2.0$ respectively. 
The long-dashed, continuous and short-dashed curves are 
for log \N(H~{\sc i}) = 20., 20.7, and 21 respectively.
 \label{figbb3e}}
\end{figure*}

Here we focus on the \h2 rotational excitation 
predicted in our calculations.

\subsubsection{The ortho-para ratio (OPR):}
The OPR indicates the kinetic temperature when the 
\h2 electronic bands are optically thick
(i.e., log(\N(\h2))$\ge$ 16; Tumlinson et al. 2002). 
Srianand et al. (2005) have shown that the OPR observed in DLAs are 
higher than those measured in Galactic ISM, LMC, and SMC sight lines. 
Here, we probe the reason for this difference.
Panel (a) of Figs.~\ref{figbb2} and  \ref{figbb2e} plots
the OPR as a function of \N(\h2) for the
stellar and diffuse cases respectively with log~\N(H~{\sc i}) = 20.7.
For optically thin \h2, when the Solomon process dominates excitations of \h2, (i.e log~\N(\h2)$\le$ 16), the predicted OPR is 
close to 3 for clouds with log~\N(H~{\sc i})=20.7. However,  
the OPR is greater than 3 for log~\N(\h2) in the range of 16 to 18. 
At \N(\h2) $\ge10^{18}~{\rm cm}^{-2}$
the OPR traces the kinetic 
temperature of the gas.

\par
Sternberg \& Neufeld (1999) show that the high value of
the OPR seen for intermediate \N(\h2) is because the electronic absorption
lines of ortho-\h2 become self-shielded at smaller column densities than  
para-\h2. Thus, ortho-\h2 
exists while para-\h2 is destroyed.
Thus, we expect the OPR to be larger in the 
case of a higher radiation field when 16$\le$log~\N(\h2)$\le$18.
For a given \N(\h2) (in the intermediate range), 
the predicted OPR is higher for higher \nh~ in models with
log~N(H~{\sc i})=20.7. 
We also notice that,
for a given \N(\h2) and \nh, the models with lower \N(H~{\sc i})
produce a lower value of the OPR (panel (a) in Fig.~\ref{figbb3e}).
%
{Thus, the observed OPR with 16$\le$log~\N(\h2)$\le$18 will
require a lower \N(H~{\sc i}) and higher \nh. This is consistent
with what we inferred based on the \N(C~{\sc i})/\N(Si~{\sc ii}) ratio.}
A detail observation of an individual system confirms that 
the components with low OPR are consistent with a low value of \N(H~{\sc i})
(see Table. 1 of Srianand et al. 2005).
As an example, \zabs = 1.96822 toward Q~0013-004 has the lowest
OPR value measured in DLAs (0.64$\pm$0.09) and has log~\N(\h2) = 16.77
and log~\N(H~{\sc i})$\le19.43$.
We notice that the kinetic temperature is in the range of $40-560$ K 
for the consistent models. This is slightly higher than the kinetic
temperature range (60$-$300K)
derived using the OPR and assuming LTE assumptions (Srianand et al. 2005).
We  notice that the OPR does not track
the kinetic temperature well in the intermediate N(\h2) range. 
A careful investigation of this is presented
elsewhere (Shaw et al. 2005).
All of our calculations are in qualitative agreement with the 
OPR seen for log~\N(\h2)$>$18 
components.

\par

\subsubsection{ N(J=4)/N(J=2) and the radiation field:}

Panel (d) of Figs.~\ref{figbb2} and \ref{figbb2e} plots {\it N(J=4)/N(J=2)} as a function of  \N(\h2) for various 
log~\N(H~{\sc i}) and $\kappa$. 
The Solomon process controls these populations since 
the energy separation between these energy 
levels is far too large and collisional excitation
is inefficient. 
This ratio indicates  $\chi$ when the \h2 column
density is low (Jura 1975). 
For a given  \N(\h2) with log  \N(\h2)$\le$ 16.0, the {\it N(J=4)/N(J=2)} ratio 
is larger for larger (i) \nh, (ii)$\kappa$ and
(iii)\N(H~{\sc i}) (panel d in Fig.~\ref{figbb3e}).
Apart from the two systems in Ledoux et al. (2003), absorption
from the {\it J}=4 level of \h2 is not detected. These
two measurements and the upper limits for the optically
thin systems are consistent with a radiation field as high as $\chi=30$.
There is very little 
difference between the diffuse and stellar continua since the excitation is mainly by electronic line absorption.
In the optically thick cases, \N(\h2) in the {\it J}=4 level is
populated mainly by formation pumping. No clear trend is present since formation pumping depends
on various quantities. 

\subsubsection{N(J=2)/N(J=0) and N(J=3)/N(J=1):}

The {\it N(J=2)/N(J=0)} ratio is more sensitive to collisional excitation than the population ratios of higher rotational levels.
The observed and predicted {\it N(J=2)/N(J=0)} and {\it N(J=3)/N(J=1)}  
are plotted as a function of \N(\h2) in 
panels (b) and (c) of Figs.~\ref{figbb2}, \ref{figbb2e}, and \ref{figbb3e} respectively.
For the intermediate range of \N(\h2) the models that reproduce 
the OPR also roughly reproduce these two ratios. However,
they do not explain the observed distribution
for log~\N(\h2)$\ge18$. 
It is important to note that the {\it J=2} and {\it J=3}
levels are mainly populated by cascades from high {\it J} levels following
formation and UV pumping.  
The grain formation distribution function and grain surface interactions
can affect the excitation of 
these high {\it J} levels.
Although fitting the observed results may shed 
light on a gas with a 
different metallicity and dust composition than the Milky Way,
such an exercise will divert us from our main theme
and is left to future work.

\subsection{\N(Al~{\sc iii})/\N(Al~{\sc ii}):}
\begin{figure}
\centerline{\epsfig{file=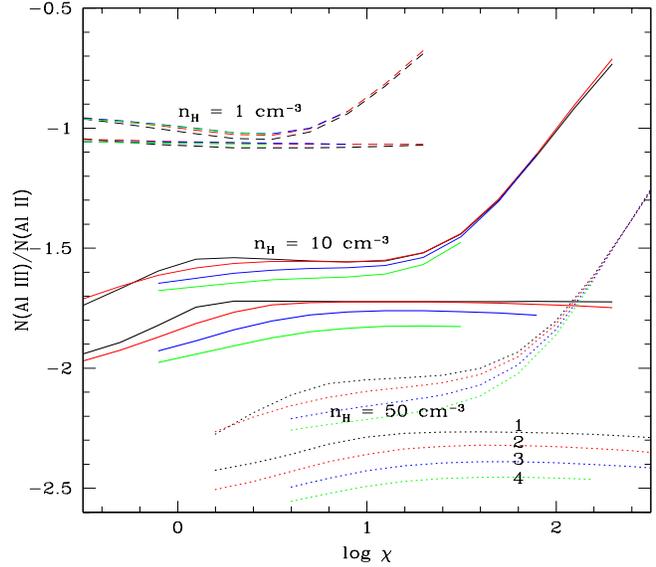,width=9cm,height=8cm}}
\caption { The ratio \N(Al~{\sc iii})/\N(Al~{\sc ii}) as a function
of $\chi$ for clouds with log~N(H~{\sc i}) = 20.7. The long-dashed,
continuous and short-dashed curves are for \nh = 1, 10, and
50 cm$^{-3}$ respectively. As in the other figures, 1, 2, 3, and 4 marks
the models with $\kappa$ $-$1.4, $-$1.6,$-$1.8 and $-$2.0 respectively.
The thin and thick curves represent the results for 
the stellar and diffuse case respectively. 
\label{bbal3}}
\end{figure}

{
Here we focus on {\N(Al~{\sc iii})/\N(Al~{\sc ii})} produced  
with a stellar radiation field on top of the Bgr(Fig.~\ref{bbal3}).  
%
Al~{\sc ii} is mainly ionized by the high energy photons from the Bgr.
%
In the diffuse case {\N(Al~{\sc iii})/\N(Al~{\sc ii})} ratio 
depends more on the density than on  $\chi$. This is
also the case for stellar case when $\chi$ is small. However, 
the ratio increases with increase in $\chi$ for large
values of $\chi$ (See thin curves in Fig.~\ref{bbal3}).
Higher $\kappa$ produces higher  
\N(Al~{\sc iii})/\N(Al~{\sc ii}), especially in the case of a 
high-density gas. The ratio  
\N(Al~{\sc iii})/\N(Al~{\sc ii}) is higher for lower \N(H~{\sc i})
for a given $\kappa$ and \nh.  
Our calculations predict log~\N(Al~{\sc iii})/\N(Al~{\sc ii}) to be 
less than $-$1.5 for the ranges of $\chi$, \nh, and $\kappa$ that 
reproduce the observed properties of the \h2 components.

\par
Srianand et al. (2005) have shown that most DLAs with log~\N(H~{\sc i}) $\ge$ 21
show C~{\sc ii$^*$} absorption even when \h2 and C~{\sc i} are
clearly absent. All these systems also show Al~{\sc iii} absorption with 
log~\N(Al~{\sc iii})/\N(Al~{\sc ii})
higher than$ -1.8$ (Table 6 of Srianand et al. 2005). 
Our model calculations produce log~\N(Al~{\sc iii})/\N(Al~{\sc ii})
higher than $-1.6$ for 
log~\N(H~{\sc i})$\ge$ 20.7 when \nh $\le$ 10 cm$^{-3}$.   
%
Clearly,  the systems with only C~{\sc ii$^*$} absorption without
\h2 and C~{\sc i} absorption need 
\nh$\le$10 cm$^{-3}$ in order to have a \N(Al~{\sc
iii})/N(Al~{\sc ii}) ratio at the detected level,
since log \N(H~{\sc i}) $\ge$21 in most of these components. 
{This range in \nh will also explain the observed 
N(C~{\sc ii$^*$})/N(Si~{\sc ii}) in these systems (see Section. 4.5).}
This
implies that C~{\sc ii$^*$} absorption originates in a 
region of lower density and higher ionization compared to the 
components that produce \h2 and C~{\sc i} absorption.

{ Lehner et al. (2004) show that a significant fraction of C~{\sc ii$*$}
absorption detected toward high latitude lines-of-sight in our
galaxy originate from warm ionized medium (WIM). Using the profile
coincidence between Al~{\sc ii}, Al~{\sc iii}, and photoionization
models (computed using Cloudy) Wolfe et al. (2004) 
argue that C~{\sc ii$^*$} in DLAs are unlikely to originate
from WIM gas. They argue that a considerable fraction of Al~{\sc iii} can
be produced from the gas that is by and large neutral
(much like the models considered here).
For the \zabs = 1.919 system toward Q~2206-19
that show C~{\sc ii$^*$} without \h2 and C~{\sc i} 
absorption lines, Wolfe et al. (2004) derive a density of
1.6 cm$^{-3}$. Srianand et al. (2005) discuss the profiles
of absorption lines from different ionization states 
(including 21 cm absorption line) in the case of \zabs = 1.944
system toward Q~1157+014 and conclude that a considerable fraction
of C~{\sc ii$^*$} absorption originate from gas at lower
densities (i.e \nh$\simeq$ 1 cm$^{-3}$). All these are consistent with our conclusion 
that density is lower (i.e \nh$\le10$ cm$^{-3}$) in the systems 
that show C~{\sc ii$*$} absorption without
\h2 and C~{\sc i} compared to the ones that show \h2
absorption (i.e \nh = 10-100 cm$^{-3}$).
}

\subsection{Other fine-structure lines:}

We predict the fine-structure level populations of O~{\sc i} and Si~{\sc ii} in addition to 
C~{\sc i} and C~{\sc ii}.  
The column densities of both O~{\sc i$^*$} and O~{\sc i$^{**}$} are in the range 2$\times10^{11} - 10^{12}$ cm$^{-2}$,
for the range of $\chi$ suggested by the C~{\sc i} and \h2 observations
(see Table.\ref{table1}). The oscillator
strengths of O~{\sc i$^*$} and O~{\sc i$^{**}$} lines are low
($\sim 4\times10^{-2}$) and 
these lines are never detected in DLAs.
Our calculations also predict
N(Si~{\sc ii$^*$}) $<$ 2$\times10^{11}$cm$^{-2}$, consistent with the fact that
Si~{\sc ii$^*$} absorption is not
detected in DLAs. It is most likely that one may not
detect these lines directly even if the absorbing gas
has a higher density due to their weakness. However, it may be possible to 
detect them by co-adding
a large number of DLA spectra.
{Thus, one possible way to confirm 
the idea that most DLAs (with or without \h2) originate in a high-density
gas with star formation is to detect excited fine-structure
lines of O I and Si II by co-adding many spectra as
one does for metal lines in the \lya
forest}.

\subsection{Summary}
The main results for a high-density cloud
in a stellar radiation field are:
\begin{itemize}
\item{} The observed properties of DLAs with \h2 are consistently reproduced 
by models with a local radiation field in addition to the QSO dominated
BGR. Most of the observations of 
the \h2 components
(such as \N(C~{\sc i})/\N(Si~{\sc ii}), \N(C~{\sc i}$^*$)/\N(C~{\sc i}) and 
\N(C~{\sc ii$^*$})/\N(Si~{\sc ii})) are consistent with lower \N(H~{\sc i})
(i,e log \N~(H~{\sc i})$\simeq$20 cm$^{-2}$) and higher densities
(10 $\le$ \nh(cm$^{-3}$) $\le$100).  
The median \N(H~{\sc i}) in DLAs with \h2 is 
$\sim$ 10$^{20.8}$ cm$^{-2}$, so in these systems only a fraction of 
the total \N(H{~\sc i}) originates in regions with \h2. The typical kinetic 
temperature ranges between 40 and 560 K.
%

\item{}We reproduce the observed range of 
the OPR in DLAs. The systems that are optically thin in the \h2 
electronic bands have a lower OPR,  
suggesting log \N(H~{\sc i})$\simeq$20, consistent with constraints 
from atomic species. The OPR $>3$,
seen in some of the components with intermediate \h2 electronic line optical
depths, are produced by the different level
of self-shielding in ortho and para \h2. 
%
The absence of 
C~{\sc i} and \h2 in {\it J} $\ge$ 4 levels in the case of a few optically
thick \h2 components are consistent with a higher $\chi$ in these clouds.

\item{} Our predictions, the measurements,
and the upper limits on the {\it N(J=4)/N(J=2)} ratio in the optically thin \h2
components are consistent with a radiation field as high as
$\chi=30$. This is consistent with the limits on the radiation field
from the atomic species.
The absence of {\it N(J=4)} lines in 
the optically thick \h2 components are consistent with a low
rate of formation pumping in these systems.

\item{} \h2 and C~{\sc i} are not detectable if the radiation
field is much higher irrespective 
of the model parameters. However, such clouds will
be easily detectable in 21 cm absorption with spin temperature
in the range of 100 to 1000 K. These clouds will also show very
strong C~{\sc ii$^*$} absorption. However, the column density
of C~{\sc ii$^*$} will strongly depend on the amount of 
ionized gas along the line of sight. Also these systems will
show very strong Al~{\sc iii} absorption.

\end{itemize}

\section{Discussion and Conclusions:}

\subsection{Nature of the radiation field:}
Ledoux et al. (2003) show that detectable \h2 absorption
(\N(H$_2$)$\ge10^{14}~{\rm cm^{-2}}$) is seen in 15-20 per cent of DLAs. 
We show that the observed properties of these systems are inconsistent
with a gas irradiated by the meta-galactic UV radiation field.
Our calculations suggest that these systems originate from a high-density gas 
($\ge10$ cm$^{-3}$) irradiated by a moderate diffuse UV radiation field 
(1 to 30 times that of Galactic ISM) and indicate ongoing star formation
in these systems. 
%
The mean
radiation field is determined by both the SFR and radiative transport.
As the mean dust optical depth in DLAs will be smaller than
that of Galactic ISM, the typical SFR in DLAs with \h2 absorption
can not be much larger than that seen in our Galaxy.
Even if such a moderate star formation exists in most DLAs,
they will still contribute appreciably to the global star formation 
rate density at higher redshifts (see Wolfe et al. 2003a,b;
Srianand et al. 2005; Hirashita \& Ferrara 2005).
\par
\subsection{Physical state of the \h2 gas:}
Our calculations with a diffuse radiation field suggest high
densities in the \h2 gas (i.e 10 $\le$ \nh(cm$^{-3}$) $\le100$).
The typical temperature of the clouds that are 
consistent with the observations are in the range 40 to
560 K. {Our calculations simultaneously explain \h2 abundance
and fine-structure excitations of atomic species
without opting for enhanced \h2 formation rate on
dust grains as required by analytical models of Hirashita
\& Ferrara (2005).
}
The inferred range in temperature and densities are 
consistent with the physical conditions in the 
CNM gas. We show that if the cloud is irradiated by a diffuse
interstellar UV background 
then \N(C~{\sc i$^*$})/\N(C~{\sc i})  
can directly probe the density of the gas. {
For radiation field with $\chi\ge10$ the ratio N(C~{\sc ii$^*$})/N(Si~{\sc ii})
will also trace the density of the gas as photo-heating saturates at higher
values of $\chi$.}
However, if the cloud is close
to the ionizing source 
 \N(C~{\sc i$^*$})/\N(C~{\sc i})
depends also on $\chi$. The predicted values of  N(C~{\sc ii$^*$})/N(Si~{\sc ii})
in the stellar case with low N(H~{\sc i}) (as required by other observations)
are much higher than the observed values. Thus, our 
calculations require that the \h2 components in DLAs are ionized by a diffuse
radiation field. {Most of the observations are consistently reproduced
with N(H~{\sc i}) = 10$^{20}$ cm$^{-2}$. This suggests that only a  
fraction of the total measured N(H~{\sc i}) is present in the \h2
components. }
%

\subsection{DLAs without \h2:}
{The observations show that systems without detectable \h2 (i.e $\sim 80-85$ per cent
of the DLAs) do not show C~{\sc i} absorption and also have very small 
values of $\kappa$. Roughly 50 per cent of the DLAs show detectable
C~{\sc ii$^*$} absorption. 
Our calculations suggest that the absence of \h2, C~{\sc i}, C~{\sc ii$^*$}
and 21 cm absorption in a considerable fraction of DLAs could just be a consequence
of a low-gas density in a  moderate radiation field  (irrespective
of dust content of the gas). This more or less agrees with the 
Wolfe et al. (2004)' conclusions that the systems with upper limits on C~{\sc ii$^*$}
absorption originate in a warm neutral medium (WNM)(also see Liszt 2002).
}

\par
{
The observed N(C~{\sc ii$^*$})/N(Si~{\sc ii}) in systems that show
detectable C~{\sc ii$^*$} absorption without \h2 and C~{\sc i} is
consistent with \nh$\ge0.1$ cm$^{-3}$ (see Section 3.3.1). 
The absence of C~{\sc i} and \h2 in these systems can be explained as a 
consequence of higher radiation field.  
Wolfe et al. (2003a; 2003b) in the framework of 
stable two phase medium argue that most of the C~{\sc ii$^*$} absorption
should originate from the CNM gas in order to have reasonable 
global star-formation rate density. 
We note \nh$<$10 cm$^{-3}$ in systems that show C~{\sc ii$^*$} without
\h2 and C~{\sc i}
so that the observed N(C~{\sc ii$^*$})/N(Si~{\sc ii}) as well as
ionization state of Al can be consistently reproduced (see Section 4.7). 
This range is consistent with the one measured by Wolfe et al. (2004)
for \zabs = 1.919 system toward 2206-10. Interestingly, the
inferred density in these systems is less than that typically required to explain
the property of \h2 detected components (i.e \nh$\ge10~cm^{-3}$). 
Thus, it appears that the systems that show only C~{\sc ii$^*$} 
seem to originate from
lower density gas compared to the ones that also show \h2 and C~{\sc i}
absorption.
}

\par
{
Unlike \h2 and C~{\sc i}, an additional radiation 
field (with h$\nu\le13.6$ eV) can not suppress 21 cm absorption in the
high-density gas. Our calculations with \nh$\ge1~{\rm cm}^{-3}$
predict a spin temperature in the range of $100-1000$ K for a range
of $\kappa$, $\chi$, and \N(H~{\sc i}) typically seen in DLAs. 
Thus, 21~cm absorption is
definitely detectable. 
Over the redshift range that is similar to the range used by
Ledoux et al. (2003) for \h2 searches (1.9$\le$\zabs$\le 3.5$),
only two out of 8 DLAs show detectable 21 cm absorption
(Kanekar \& Chengalur 2003). The rest of these systems have 
a lower limit on the spin temperature in the range of 700-9000 K.
Both of the systems with 21 cm absorption
also show detectable C~{\sc ii$^*$} absorption. 
Detail investigation of one of these systems (\zabs = 1.944 system toward 
Q 1157+014) shows that C~{\sc ii$^*$} originate not only from 
the gas responsible for 21 cm absorption but also from other 
components (Fig. 19 Srianand et al 2005). Clearly C~{\sc ii$^*$}
traces a wider range of physical conditions.
There are few systems
(e.g \zabs = 3.387 toward Q~0201+11 and \zabs = 3.063 toward Q 0336-01)
that show detectable C~{\sc ii$^*$} absorption without 21
cm absorption. The derived upper limits on spin temperatures in these
systems will mean very low CNM fraction along the sight lines if
the gas covers the background radio source completely (Kanekar \& Chengalur, 2003). 
In the absence of
VLBI observations, interpretation of these system will be very
subjective (see Wolfe et al. 2003b for details). 
A careful analysis of Al~{\sc iii}/Al~{\sc ii} and 
N(C~{\sc ii$^*$})/N(Si~{\sc ii}) in individual components is needed
to get the contribution of ionized gas to the excitations of C~{\sc ii$^*$}.

\par
Alternatively, one can use the fine-structure state populations of 
O~{\sc i} and Si~{\sc ii} that trace C~{\sc ii} very closely.
Our calculations also compute expected fine-structure excitations of O~{\sc i}
and Si~{\sc ii}. The expected column densities of O~{\sc i$^*$}, O~{\sc i$^{**}$}
and Si~{\sc ii$^*$} are in the range $10^{11}-10^{12}$ cm$^{-2}$ for the range of
parameters considered in our calculations. It may be possible to detect these
lines using pixel optical depth techniques that are used to detect
metals in the diffuse low density IGM. 
Detection of such lines will put stringent constraints
on density in these systems.
}
\par
\subsection{Conclusions:}
{
In this article, we present calculations that self-consistently determine the 
gas ionization, level populations (atomic fine-structure levels and
rotational levels of \h2), grain physics, and chemistry. 
We show that 
for a low-density gas (\nh$\le$ 0.1 cm$^{-3}$) the meta-galactic 
UV background due to quasars is sufficient to maintain  \h2 
column densities below the detection limit (i.e \N(\h2)$\le10^{14}$ cm$^{-2}$) 
irrespective of the metallicity and dust content in the gas. Such a gas will have
a 21 cm spin temperature in excess of 7000 K and very low C~{\sc i}
and C~{\sc ii$^*$} column densities for H~{\sc i} column densities
typically observed in DLAs. 

\par
Calculations with a high-density gas in the presence of a local radiation field reproduce
most of the observations of \h2 components in DLAs.
Thus our study clearly confirms the presence of CNM at least in 15-20\%
of the DLAs. We also show only fraction of total N(H~{\sc i}) is in
the \h2 components. 

\par
Unlike the components with \h2, interpretation of systems that show only C~{\sc ii$^*$}
without additional constraints is not clear. 
This is because presence of free electrons
can be more efficient in populating the fine-structure level of
C~{\sc ii}. This can lead to a high value of inferred \nh~if 
the electron contribution is neglected.
Using Al~{\sc iii}
absorption we show that a gas that produces  C~{\sc ii$^*$} in 
systems without \h2 has lower density than the ones with \h2 absorption.
}
%
%
%
%
%
%
\par\noindent
\section{acknowledgements} GJF acknowledges the support of the NSF 
through AST 03-07720 and NASA with grant NAG5-65692. 
GJF and RS acknowledge the support from the DST/INT/US(NSF-RP0-115)/2002.
GS would like to thank CCS, University of Kentucky for their two years of support.
The hospitality of 
IUCAA is gratefully acknowledged. RS and PPJ gratefully acknowledge support 
from the Indo-French
Centre for the Promotion of Advanced Research (Centre Franco-Indien pour
la Promotion de la Recherche Avanc\'ee) under contract No. 3004-3.

\end{document}